\newif\ifANON{}
\newif\ifACM{}
\newif\ifELSEVIER{}
\newif\ifLNCS{}
\newif\ifIACRTRANS{}
\newif\ifIEEE{}
\newif\ifUSENIX{}
\newif\ifSREP{}
\newif\ifLINUXBUILD{}

\ELSEVIERtrue{} %

\IfFileExists{.tmp.anon.tmp}{\ANONtrue}{}
\IfFileExists{/dev/null}{\LINUXBUILDtrue}{}

\ifELSEVIER{}
\PassOptionsToPackage{table}{xcolor}
\documentclass[5p,times,sort]{elsarticle}
\fi

\usepackage[T1]{fontenc}
\usepackage{graphicx}
\usepackage[final]{microtype}
\usepackage{xspace}
\usepackage{lipsum}
\usepackage{multirow}
\usepackage{array} %
\usepackage{booktabs} %
\usepackage{hyperref} %
\usepackage{url} %
\ifELSEVIER{}
\usepackage{xurl} %
\fi
\usepackage{array} %
\usepackage{placeins} %
\usepackage{subcaption} %
\usepackage[shortcuts]{glossaries} %
\usepackage{listings}
\usepackage{xcolor}

\usepackage{amsthm}
\newtheorem{takeout}{\textbf{Takeaway}}  %
\newtheorem{lesson}{\textbf{Lesson}}
\glsdisablehyper{}

\usepackage{todonotes}

\usepackage{xcolor}
\usepackage{tcolorbox}
\usepackage[inline]{enumitem}
\lstdefinelanguage{PowerShell}{
    morekeywords={
            Add-Content,Add-PSSnapin,Clear-Content,Clear-History,Clear-Host,Clear-Item,Clear-ItemProperty,Clear-Variable,Compare-Object,Connect-PSSession,ConvertFrom-String,Convert-Path,Copy-Item,Copy-ItemProperty,Disable-PSBreakpoint,Disconnect-PSSession,Enable-PSBreakpoint,Enter-PSSession,Exit-PSSession,Export-Alias,Export-Csv,Export-PSSession,ForEach-Object,Format-Custom,Format-Hex,Format-List,Format-Table,Format-Wide,Get-Alias,Get-ChildItem,Get-Clipboard,Get-Command,Get-ComputerInfo,Get-Content,Get-History,Get-Item,Get-ItemProperty,Get-ItemPropertyValue,Get-Job,Get-Location,Get-Member,Get-Module,Get-Process,Get-PSBreakpoint,Get-PSCallStack,Get-PSDrive,Get-PSSession,Get-PSSnapin,Get-Service,Get-TimeZone,Get-Unique,Get-Variable,Get-WmiObject,Group-Object,help,Import-Alias,Import-Csv,Import-Module,Import-PSSession,Invoke-Command,Invoke-Expression,Invoke-History,Invoke-Item,Invoke-RestMethod,Invoke-WebRequest,Invoke-WmiMethod,Measure-Object,mkdir,Move-Item,Move-ItemProperty,New-Alias,New-Item,New-Module,New-PSDrive,New-PSSession,New-PSSessionConfigurationFile,New-Variable,Out-GridView,Out-Host,Out-Printer,Pop-Location,powershell_ise.exe,Push-Location,Receive-Job,Receive-PSSession,Remove-Item,Remove-ItemProperty,Remove-Job,Remove-Module,Remove-PSBreakpoint,Remove-PSDrive,Remove-PSSession,Remove-PSSnapin,Remove-Variable,Remove-WmiObject,Rename-Item,Rename-ItemProperty,Resolve-Path,Resume-Job,Select-Object,Select-String,Set-Alias,Set-Clipboard,Set-Content,Set-Item,Set-ItemProperty,Set-Location,Set-PSBreakpoint,Set-TimeZone,Set-Variable,Set-WmiInstance,Show-Command,Sort-Object,Start-Job,Start-Process,Start-Service,Start-Sleep,Stop-Job,Stop-Process,Stop-Service,Suspend-Job,Tee-Object,Trace-Command,Wait-Job,Where-Object,Write-Output
    },
    morekeywords={
        Add-AppxPackage,Add-AppxProvisionedPackage,Add-AppxVolume,Add-BitsFile,Add-CertificateEnrollmentPolicyServer,Add-Computer,Add-Content,Add-History,Add-JobTrigger,Add-KdsRootKey,Add-LocalGroupMember,Add-Member,Add-PSSnapin,Add-Type,Add-WindowsCapability,Add-WindowsDriver,Add-WindowsImage,Add-WindowsPackage,Checkpoint-Computer,Clear-Content,Clear-EventLog,Clear-History,Clear-Item,Clear-ItemProperty,Clear-KdsCache,Clear-RecycleBin,Clear-Tpm,Clear-Variable,Clear-WindowsCorruptMountPoint,Compare-Object,Complete-BitsTransfer,Complete-DtiagnosticTransaction,Complete-Transaction,Confirm-SecureBootUEFI,Connect-PSSession,Connect-WSMan,ConvertFrom-Csv,ConvertFrom-Json,ConvertFrom-SecureString,ConvertFrom-String,ConvertFrom-StringData,Convert-Path,Convert-String,ConvertTo-Csv,ConvertTo-Html,ConvertTo-Json,ConvertTo-ProcessMitigationPolicy,ConvertTo-SecureString,ConvertTo-TpmOwnerAuth,ConvertTo-Xml,Copy-Item,Copy-ItemProperty,Debug-Job,Debug-Process,Debug-Runspace,Disable-AppBackgroundTaskDiagnosticLog,Disable-ComputerRestore,Disable-JobTrigger,Disable-LocalUser,Disable-PSBreakpoint,Disable-PSRemoting,Disable-PSSessionConfiguration,Disable-RunspaceDebug,Disable-ScheduledJob,Disable-TlsCipherSuite,Disable-TlsEccCurve,Disable-TlsSessionTicketKey,Disable-TpmAutoProvisioning,Disable-WindowsErrorReporting,Disable-WindowsOptionalFeature,Disable-WSManCredSSP,Disconnect-PSSession,Disconnect-WSMan,Dismount-AppxVolume,Dismount-WindowsImage,Enable-AppBackgroundTaskDiagnosticLog,Enable-ComputerRestore,Enable-JobTrigger,Enable-LocalUser,Enable-PSBreakpoint,Enable-PSRemoting,Enable-PSSessionConfiguration,Enable-RunspaceDebug,Enable-ScheduledJob,Enable-TlsCipherSuite,Enable-TlsEccCurve,Enable-TlsSessionTicketKey,Enable-TpmAutoProvisioning,Enable-WindowsErrorReporting,Enable-WindowsOptionalFeature,Enable-WSManCredSSP,Enter-PSHostProcess,Enter-PSSession,Exit-PSHostProcess,Exit-PSSession,Expand-WindowsCustomDataImage,Expand-WindowsImage,Export-Alias,Export-BinaryMiLog,Export-Certificate,Export-Clixml,Export-Console,Export-Counter,Export-Csv,Export-FormatData,Export-ModuleMember,Export-PfxCertificate,Export-ProvisioningPackage,Export-PSSession,Export-StartLayout,Export-StartLayoutEdgeAssets,Export-TlsSessionTicketKey,Export-Trace,Export-WindowsCapabilitySource,Export-WindowsDriver,Export-WindowsImage,Find-Package,Find-PackageProvider,ForEach-Object,Format-Custom,Format-List,Format-SecureBootUEFI,Format-Table,Format-Wide,Get-Acl,Get-Alias,Get-AppxDefaultVolume,Get-AppxPackage,Get-AppxPackageManifest,Get-AppxProvisionedPackage,Get-AppxVolume,Get-AuthenticodeSignature,Get-BitsTransfer,Get-Certificate,Get-CertificateAutoEnrollmentPolicy,Get-CertificateEnrollmentPolicyServer,Get-CertificateNotificationTask,Get-ChildItem,Get-CimAssociatedInstance,Get-CimClass,Get-CimInstance,Get-CimSession,Get-Clipboard,Get-CmsMessage,Get-Command,Get-ComputerInfo,Get-ComputerRestorePoint,Get-Content,Get-ControlPanelItem,Get-Counter,Get-Credential,Get-Culture,Get-DAPolicyChange,Get-Date,Get-DeliveryOptimizationLog,Get-DeliveryOptimizationPerfSnap,Get-DeliveryOptimizationPerfSnapThisMonth,Get-DeliveryOptimizationStatus,Get-DODownloadMode,Get-DOPercentageMaxBackgroundBandwidth,Get-DOPercentageMaxForegroundBandwidth,Get-Event,Get-EventLog,Get-EventSubscriber,Get-ExecutionPolicy,Get-FormatData,Get-Help,Get-History,Get-Host,Get-HotFix,Get-Item,Get-ItemProperty,Get-ItemPropertyValue,Get-Job,Get-JobTrigger,Get-KdsConfiguration,Get-KdsRootKey,Get-LocalGroup,Get-LocalGroupMember,Get-LocalUser,Get-Location,Get-Member,Get-Module,Get-Package,Get-PackageProvider,Get-PackageSource,Get-PfxCertificate,Get-PfxData,Get-PmemDisk,Get-PmemPhysicalDevice,Get-PmemUnusedRegion,Get-Process,Get-ProcessMitigation,Get-ProvisioningPackage,Get-PSBreakpoint,Get-PSCallStack,Get-PSDrive,Get-PSHostProcessInfo,Get-PSProvider,Get-PSReadlineKeyHandler,Get-PSReadlineOption,Get-PSSession,Get-PSSessionCapability,Get-PSSessionConfiguration,Get-PSSnapin,Get-Random,Get-Runspace,Get-RunspaceDebug,Get-ScheduledJob,Get-ScheduledJobOption,Get-SecureBootPolicy,Get-SecureBootUEFI,Get-Service,Get-TimeZone,Get-TlsCipherSuite,Get-TlsEccCurve,Get-Tpm,Get-TpmEndorsementKeyInfo,Get-TpmSupportedFeature,Get-TraceSource,Get-Transaction,Get-TroubleshootingPack,Get-TrustedProvisioningCertificate,Get-TypeData,Get-UICulture,Get-Unique,Get-Variable,Get-WIMBootEntry,Get-WinAcceptLanguageFromLanguageListOptOut,Get-WinCultureFromLanguageListOptOut,Get-WinDefaultInputMethodOverride,Get-WindowsCapability,Get-WindowsDeveloperLicense,Get-WindowsDriver,Get-WindowsEdition,Get-WindowsErrorReporting,Get-WindowsImage,Get-WindowsImageContent,Get-WindowsOptionalFeature,Get-WindowsPackage,Get-WindowsSearchSetting,Get-WinEvent,Get-WinHomeLocation,Get-WinLanguageBarOption,Get-WinSystemLocale,Get-WinUILanguageOverride,Get-WinUserLanguageList,Get-WmiObject,Get-WSManCredSSP,Get-WSManInstance,Group-Object,Import-Alias,Import-BinaryMiLog,Import-Certificate,Import-Clixml,Import-Counter,Import-Csv,Import-LocalizedData,Import-Module,Import-PackageProvider,Import-PfxCertificate,Import-PSSession,Import-StartLayout,Import-TpmOwnerAuth,Initialize-PmemPhysicalDevice,Initialize-Tpm,Install-Package,Install-PackageProvider,Install-ProvisioningPackage,Install-TrustedProvisioningCertificate,Invoke-CimMethod,Invoke-Command,Invoke-CommandInDesktopPackage,Invoke-DscResource,Invoke-Expression,Invoke-History,Invoke-Item,Invoke-RestMethod,Invoke-TroubleshootingPack,Invoke-WebRequest,Invoke-WmiMethod,Invoke-WSManAction,Join-DtiagnosticResourceManager,Join-Path,Limit-EventLog,Measure-Command,Measure-Object,Mount-AppxVolume,Mount-WindowsImage,Move-AppxPackage,Move-Item,Move-ItemProperty,New-Alias,New-CertificateNotificationTask,New-CimInstance,New-CimSession,New-CimSessionOption,New-DtiagnosticTransaction,New-Event,New-EventLog,New-FileCatalog,New-Item,New-ItemProperty,New-JobTrigger,New-LocalGroup,New-LocalUser,New-Module,New-ModuleManifest,New-NetIPsecAuthProposal,New-NetIPsecMainModeCryptoProposal,New-NetIPsecQuickModeCryptoProposal,New-Object,New-PmemDisk,New-ProvisioningRepro,New-PSDrive,New-PSRoleCapabilityFile,New-PSSession,New-PSSessionConfigurationFile,New-PSSessionOption,New-PSTransportOption,New-PSWorkflowExecutionOption,New-ScheduledJobOption,New-SelfSignedCertificate,New-Service,New-TimeSpan,New-TlsSessionTicketKey,New-Variable,New-WebServiceProxy,New-WindowsCustomImage,New-WindowsImage,New-WinEvent,New-WinUserLanguageList,New-WSManInstance,New-WSManSessionOption,Optimize-AppxProvisionedPackages,Optimize-WindowsImage,Out-Default,Out-File,Out-GridView,Out-Host,Out-Null,Out-Printer,Out-String,Pop-Location,Protect-CmsMessage,Publish-DscConfiguration,Push-Location,Read-Host,Receive-DtiagnosticTransaction,Receive-Job,Receive-PSSession,Register-ArgumentCompleter,Register-CimIndicationEvent,Register-EngineEvent,Register-ObjectEvent,Register-PackageSource,Register-PSSessionConfiguration,Register-ScheduledJob,Register-WmiEvent,Remove-AppxPackage,Remove-AppxProvisionedPackage,Remove-AppxVolume,Remove-BitsTransfer,Remove-CertificateEnrollmentPolicyServer,Remove-CertificateNotificationTask,Remove-CimInstance,Remove-CimSession,Remove-Computer,Remove-Event,Remove-EventLog,Remove-Item,Remove-ItemProperty,Remove-Job,Remove-JobTrigger,Remove-LocalGroup,Remove-LocalGroupMember,Remove-LocalUser,Remove-Module,Remove-PmemDisk,Remove-PSBreakpoint,Remove-PSDrive,Remove-PSReadlineKeyHandler,Remove-PSSession,Remove-PSSnapin,Remove-TypeData,Remove-Variable,Remove-WindowsCapability,Remove-WindowsDriver,Remove-WindowsImage,Remove-WindowsPackage,Remove-WmiObject,Remove-WSManInstance,Rename-Computer,Rename-Item,Rename-ItemProperty,Rename-LocalGroup,Rename-LocalUser,Repair-WindowsImage,Reset-ComputerMachinePassword,Resolve-DnsName,Resolve-Path,Restart-Computer,Restart-Service,Restore-Computer,Resume-BitsTransfer,Resume-Job,Resume-ProvisioningSession,Resume-Service,Save-Help,Save-Package,Save-WindowsImage,Select-Object,Select-String,Select-Xml,Send-DtiagnosticTransaction,Send-MailMessage,Set-Acl,Set-Alias,Set-AppBackgroundTaskResourcePolicy,Set-AppxDefaultVolume,Set-AppXProvisionedDataFile,Set-AuthenticodeSignature,Set-BitsTransfer,Set-CertificateAutoEnrollmentPolicy,Set-CimInstance,Set-Clipboard,Set-Content,Set-Culture,Set-Date,Set-DODownloadMode,Set-DOPercentageMaxBackgroundBandwidth,Set-DOPercentageMaxForegroundBandwidth,Set-DscLocalConfigurationManager,Set-ExecutionPolicy,Set-Item,Set-ItemProperty,Set-JobTrigger,Set-KdsConfiguration,Set-LocalGroup,Set-LocalUser,Set-Location,Set-PackageSource,Set-ProcessMitigation,Set-PSBreakpoint,Set-PSDebug,Set-PSReadlineKeyHandler,Set-PSReadlineOption,Set-PSSessionConfiguration,Set-ScheduledJob,Set-ScheduledJobOption,Set-SecureBootUEFI,Set-Service,Set-StrictMode,Set-TimeZone,Set-TpmOwnerAuth,Set-TraceSource,Set-Variable,Set-WinAcceptLanguageFromLanguageListOptOut,Set-WinCultureFromLanguageListOptOut,Set-WinDefaultInputMethodOverride,Set-WindowsEdition,Set-WindowsProductKey,Set-WindowsSearchSetting,Set-WinHomeLocation,Set-WinLanguageBarOption,Set-WinSystemLocale,Set-WinUILanguageOverride,Set-WinUserLanguageList,Set-WmiInstance,Set-WSManInstance,Set-WSManQuickConfig,Show-Command,Show-ControlPanelItem,Show-EventLog,Show-WindowsDeveloperLicenseRegistration,Sort-Object,Split-Path,Split-WindowsImage,Start-BitsTransfer,Start-DscConfiguration,Start-DtiagnosticResourceManager,Start-Job,Start-Process,Start-Service,Start-Sleep,Start-Transaction,Start-Transcript,Stop-Computer,Stop-DtiagnosticResourceManager,Stop-Job,Stop-Process,Stop-Service,Stop-Transcript,Suspend-BitsTransfer,Suspend-Job,Suspend-Service,Switch-Certificate,Tee-Object,Test-Certificate,Test-ComputerSecureChannel,Test-Connection,Test-DscConfiguration,Test-FileCatalog,Test-KdsRootKey,Test-ModuleManifest,Test-Path,Test-PSSessionConfigurationFile,Test-WSMan,Trace-Command,Unblock-File,Unblock-Tpm,Undo-DtiagnosticTransaction,Undo-Transaction,Uninstall-Package,Uninstall-ProvisioningPackage,Uninstall-TrustedProvisioningCertificate,Unprotect-CmsMessage,Unregister-Event,Unregister-PackageSource,Unregister-PSSessionConfiguration,Unregister-ScheduledJob,Unregister-WindowsDeveloperLicense,Update-FormatData,Update-Help,Update-List,Update-TypeData,Update-WIMBootEntry,Use-Transaction,Use-WindowsUnattend,Wait-Debugger,Wait-Event,Wait-Job,Wait-Process,Where-Object,Write-Debug,Write-Error,Write-EventLog,Write-Host,Write-Information,Write-Output,Write-Progress,Write-Verbose,Write-Warning
    },
    morekeywords={
        Add-BitLockerKeyProtector,Add-DnsClientNrptRule,Add-DtcClusterTMMapping,Add-EtwTraceProvider,Add-InitiatorIdToMaskingSet,Add-MpPreference,Add-NetEventNetworkAdapter,Add-NetEventPacketCaptureProvider,Add-NetEventProvider,Add-NetEventVFPProvider,Add-NetEventVmNetworkAdapter,Add-NetEventVmSwitch,Add-NetEventVmSwitchProvider,Add-NetEventWFPCaptureProvider,Add-NetIPHttpsCertBinding,Add-NetLbfoTeamMember,Add-NetLbfoTeamNic,Add-NetNatExternalAddress,Add-NetNatStaticMapping,Add-NetSwitchTeamMember,Add-Odbsn,Add-PartitionAccessPath,Add-PhysicalDisk,Add-Printer,Add-PrinterDriver,Add-PrinterPort,Add-StorageFaultDomain,Add-TargetPortToMaskingSet,Add-VirtualDiskToMaskingSet,Add-VpnConnection,Add-VpnConnectionRoute,Add-VpnConnectionTriggerApplication,Add-VpnConnectionTriggerDnsConfiguration,Add-VpnConnectionTriggerTrustedNetwork,AfterAll,AfterEach,Assert-MockCalled,Assert-VerifiableMocks,Backup-BitLockerKeyProtector,BackupToAAD-BitLockerKeyProtector,BeforeAll,BeforeEach,Block-FileShareAccess,Block-SmbShareAccess,Clear-BitLockerAutoUnlock,Clear-Disk,Clear-DnsClientCache,Clear-FileStorageTier,Clear-Host,Clear-PcsvDeviceLog,Clear-StorageDiagnosticInfo,Close-SmbOpenFile,Close-SmbSession,Compress-Archive,Configuration,Connect-IscsiTarget,Connect-VirtualDisk,Context,convert,ConvertFrom-SddlString,Copy-NetFirewallRule,Copy-NetIPsecMainModeCryptoSet,Copy-NetIPsecMainModeRule,Copy-NetIPsecPhase1AuthSet,Copy-NetIPsecPhase2AuthSet,Copy-NetIPsecQuickModeCryptoSet,Copy-NetIPsecRule,Debug-FileShare,Debug-MMAppPrelaunch,Debug-StorageSubSystem,Debug-Volume,Describe,Disable-BitLocker,Disable-BitLockerAutoUnlock,Disable-DAManualEntryPointSelection,Disable-Dsebug,Disable-MMAgent,Disable-NetAdapter,Disable-NetAdapterBinding,Disable-NetAdapterChecksumOffload,Disable-NetAdapterEncapsulatedPacketTaskOffload,Disable-NetAdapterIPsecOffload,Disable-NetAdapterLso,Disable-NetAdapterPacketDirect,Disable-NetAdapterPowerManagement,Disable-NetAdapterQos,Disable-NetAdapterRdma,Disable-NetAdapterRsc,Disable-NetAdapterRss,Disable-NetAdapterSriov,Disable-NetAdapterVmq,Disable-NetDnsTransitionConfiguration,Disable-NetFirewallRule,Disable-NetIPHttpsProfile,Disable-NetIPsecMainModeRule,Disable-NetIPsecRule,Disable-NetNatTransitionConfiguration,Disable-NetworkSwitchEthernetPort,Disable-NetworkSwitchFeature,Disable-NetworkSwitchVlan,Disable-OdbcPerfCounter,Disable-PhysicalDiskIdentification,Disable-PnpDevice,Disable-PSTrace,Disable-PSWSManCombinedTrace,Disable-ScheduledTask,Disable-SmbDelegation,Disable-StorageEnclosureIdentification,Disable-StorageEnclosurePower,Disable-StorageHighAvailability,Disable-StorageMaintenanceMode,Disable-WdacBidTrace,Disable-WSManTrace,Disconnect-IscsiTarget,Disconnect-VirtualDisk,Dismount-DiskImage,Enable-BitLocker,Enable-BitLockerAutoUnlock,Enable-DAManualEntryPointSelection,Enable-Dsebug,Enable-MMAgent,Enable-NetAdapter,Enable-NetAdapterBinding,Enable-NetAdapterChecksumOffload,Enable-NetAdapterEncapsulatedPacketTaskOffload,Enable-NetAdapterIPsecOffload,Enable-NetAdapterLso,Enable-NetAdapterPacketDirect,Enable-NetAdapterPowerManagement,Enable-NetAdapterQos,Enable-NetAdapterRdma,Enable-NetAdapterRsc,Enable-NetAdapterRss,Enable-NetAdapterSriov,Enable-NetAdapterVmq,Enable-NetDnsTransitionConfiguration,Enable-NetFirewallRule,Enable-NetIPHttpsProfile,Enable-NetIPsecMainModeRule,Enable-NetIPsecRule,Enable-NetNatTransitionConfiguration,Enable-NetworkSwitchEthernetPort,Enable-NetworkSwitchFeature,Enable-NetworkSwitchVlan,Enable-OdbcPerfCounter,Enable-PhysicalDiskIdentification,Enable-PnpDevice,Enable-PSTrace,Enable-PSWSManCombinedTrace,Enable-ScheduledTask,Enable-SmbDelegation,Enable-StorageEnclosureIdentification,Enable-StorageEnclosurePower,Enable-StorageHighAvailability,Enable-StorageMaintenanceMode,Enable-WdacBidTrace,Enable-WSManTrace,Expand-Archive,Export-ODataEndpointProxy,Export-ScheduledTask,Find-Command,Find-DscResource,Find-Module,Find-NetIPsecRule,Find-NetRoute,Find-RoleCapability,Find-Script,Flush-EtwTraceSession,Format-Hex,Format-Volume,Get-AppBackgroundTask,Get-AppxLastError,Get-AppxLog,Get-AutologgerConfig,Get-BitLockerVolume,Get-ClusteredScheduledTask,Get-DAClientExperienceConfiguration,Get-DAConnectionStatus,Get-DAEntryPointTableItem,Get-DedupProperties,Get-Disk,Get-DiskImage,Get-DiskStorageNodeView,Get-DnsClient,Get-DnsClientCache,Get-DnsClientGlobalSetting,Get-DnsClientNrptGlobal,Get-DnsClientNrptPolicy,Get-DnsClientNrptRule,Get-DnsClientServerAddress,Get-DscConfiguration,Get-DscConfigurationStatus,Get-DscLocalConfigurationManager,Get-DscResource,Get-Dtc,Get-DtcAdvancedHostSetting,Get-DtcAdvancedSetting,Get-DtcClusterDefault,Get-DtcClusterTMMapping,Get-Dtefault,Get-DtcLog,Get-DtcNetworkSetting,Get-DtcTransaction,Get-DtcTransactionsStatistics,Get-DtcTransactionsTraceSession,Get-DtcTransactionsTraceSetting,Get-EtwTraceProvider,Get-EtwTraceSession,Get-FileHash,Get-FileIntegrity,Get-FileShare,Get-FileShareAccessControlEntry,Get-FileStorageTier,Get-InitiatorId,Get-InitiatorPort,Get-InstalledModule,Get-InstalledScript,Get-IscsiConnection,Get-IscsiSession,Get-IscsiTarget,Get-IscsiTargetPortal,Get-IseSnippet,Get-LogProperties,Get-MaskingSet,Get-MMAgent,Get-MockDynamicParameters,Get-MpComputerStatus,Get-MpPreference,Get-MpThreat,Get-MpThreatCatalog,Get-MpThreatDetection,Get-NCSIPolicyConfiguration,Get-Net6to4Configuration,Get-NetAdapter,Get-NetAdapterAdvancedProperty,Get-NetAdapterBinding,Get-NetAdapterChecksumOffload,Get-NetAdapterEncapsulatedPacketTaskOffload,Get-NetAdapterHardwareInfo,Get-NetAdapterIPsecOffload,Get-NetAdapterLso,Get-NetAdapterPacketDirect,Get-NetAdapterPowerManagement,Get-NetAdapterQos,Get-NetAdapterRdma,Get-NetAdapterRsc,Get-NetAdapterRss,Get-NetAdapterSriov,Get-NetAdapterSriovVf,Get-NetAdapterStatistics,Get-NetAdapterVmq,Get-NetAdapterVMQQueue,Get-NetAdapterVPort,Get-NetCompartment,Get-NetConnectionProfile,Get-NetDnsTransitionConfiguration,Get-NetDnsTransitionMonitoring,Get-NetEventNetworkAdapter,Get-NetEventPacketCaptureProvider,Get-NetEventProvider,Get-NetEventSession,Get-NetEventVFPProvider,Get-NetEventVmNetworkAdapter,Get-NetEventVmSwitch,Get-NetEventVmSwitchProvider,Get-NetEventWFPCaptureProvider,Get-NetFirewallAddressFilter,Get-NetFirewallApplicationFilter,Get-NetFirewallInterfaceFilter,Get-NetFirewallInterfaceTypeFilter,Get-NetFirewallPortFilter,Get-NetFirewallProfile,Get-NetFirewallRule,Get-NetFirewallSecurityFilter,Get-NetFirewallServiceFilter,Get-NetFirewallSetting,Get-NetIPAddress,Get-NetIPConfiguration,Get-NetIPHttpsConfiguration,Get-NetIPHttpsState,Get-NetIPInterface,Get-NetIPseospSetting,Get-NetIPsecMainModeCryptoSet,Get-NetIPsecMainModeRule,Get-NetIPsecMainModeSA,Get-NetIPsecPhase1AuthSet,Get-NetIPsecPhase2AuthSet,Get-NetIPsecQuickModeCryptoSet,Get-NetIPsecQuickModeSA,Get-NetIPsecRule,Get-NetIPv4Protocol,Get-NetIPv6Protocol,Get-NetIsatapConfiguration,Get-NetLbfoTeam,Get-NetLbfoTeamMember,Get-NetLbfoTeamNic,Get-NetNat,Get-NetNatExternalAddress,Get-NetNatGlobal,Get-NetNatSession,Get-NetNatStaticMapping,Get-NetNatTransitionConfiguration,Get-NetNatTransitionMonitoring,Get-NetNeighbor,Get-NetOffloadGlobalSetting,Get-NetPrefixPolicy,Get-NetQosPolicy,Get-NetRoute,Get-NetSwitchTeam,Get-NetSwitchTeamMember,Get-NetTCPConnection,Get-NetTCPSetting,Get-NetTeredoConfiguration,Get-NetTeredoState,Get-NetTransportFilter,Get-NetUDPEndpoint,Get-NetUDPSetting,Get-NetworkSwitchEthernetPort,Get-NetworkSwitchFeature,Get-NetworkSwitchGlobalData,Get-NetworkSwitchVlan,Get-Odbriver,Get-Odbsn,Get-OdbcPerfCounter,Get-OffloadDataTransferSetting,Get-OperationValidation,Get-Partition,Get-PartitionSupportedSize,Get-PcsvDevice,Get-PcsvDeviceLog,Get-PhysicalDisk,Get-PhysicalDiskStorageNodeView,Get-PhysicalExtent,Get-PhysicalExtentAssociation,Get-PnpDevice,Get-PnpDeviceProperty,Get-PrintConfiguration,Get-Printer,Get-PrinterDriver,Get-PrinterPort,Get-PrinterProperty,Get-PrintJob,Get-PSRepository,Get-ResiliencySetting,Get-ScheduledTask,Get-ScheduledTaskInfo,Get-SmbBandWidthLimit,Get-SmbClientConfiguration,Get-SmbClientNetworkInterface,Get-SmbConnection,Get-SmbDelegation,Get-SmbGlobalMapping,Get-SmbMapping,Get-SmbMultichannelConnection,Get-SmbMultichannelConstraint,Get-SmbOpenFile,Get-SmbServerConfiguration,Get-SmbServerNetworkInterface,Get-SmbSession,Get-SmbShare,Get-SmbShareAccess,Get-SmbWitnessClient,Get-StartApps,Get-StorageAdvancedProperty,Get-StorageDiagnosticInfo,Get-StorageEnclosure,Get-StorageEnclosureStorageNodeView,Get-StorageEnclosureVendorData,Get-StorageExtendedStatus,Get-StorageFaultDomain,Get-StorageFileServer,Get-StorageFirmwareInformation,Get-StorageHealthAction,Get-StorageHealthReport,Get-StorageHealthSetting,Get-StorageJob,Get-StorageNode,Get-StoragePool,Get-StorageProvider,Get-StorageReliabilityCounter,Get-StorageSetting,Get-StorageSubSystem,Get-StorageTier,Get-StorageTierSupportedSize,Get-SupportedClusterSizes,Get-SupportedFileSystems,Get-TargetPort,Get-TargetPortal,Get-TestDriveItem,Get-Verb,Get-VirtualDisk,Get-VirtualDiskSupportedSize,Get-Volume,Get-VolumeCorruptionCount,Get-VolumeScrubPolicy,Get-VpnConnection,Get-VpnConnectionTrigger,Get-WdacBidTrace,Get-WindowsUpdateLog,Get-WUAVersion,Get-WUIsPendingReboot,Get-WULastInstallationDate,Get-WULastScanSuccessDate,Grant-FileShareAccess,Grant-SmbShareAccess,help,Hide-VirtualDisk,Import-IseSnippet,Import-PowerShellDataFile,ImportSystemModules,In,Initialize-Disk,InModuleScope,Install-Dtc,Install-Module,Install-Script,Install-WUUpdates,Invoke-AsWorkflow,Invoke-Mock,Invoke-OperationValidation,Invoke-Pester,It,Lock-BitLocker,mkdir,Mock,more,Mount-DiskImage,Move-SmbWitnessClient,New-AutologgerConfig,New-DAEntryPointTableItem,New-DscChecksum,New-EapConfiguration,New-EtwTraceSession,New-FileShare,New-Fixture,New-Guid,New-IscsiTargetPortal,New-IseSnippet,New-MaskingSet,New-NetAdapterAdvancedProperty,New-NetEventSession,New-NetFirewallRule,New-NetIPAddress,New-NetIPHttpsConfiguration,New-NetIPseospSetting,New-NetIPsecMainModeCryptoSet,New-NetIPsecMainModeRule,New-NetIPsecPhase1AuthSet,New-NetIPsecPhase2AuthSet,New-NetIPsecQuickModeCryptoSet,New-NetIPsecRule,New-NetLbfoTeam,New-NetNat,New-NetNatTransitionConfiguration,New-NetNeighbor,New-NetQosPolicy,New-NetRoute,New-NetSwitchTeam,New-NetTransportFilter,New-NetworkSwitchVlan,New-Partition,New-PesterOption,New-PSWorkflowSession,New-ScheduledTask,New-ScheduledTaskAction,New-ScheduledTaskPrincipal,New-ScheduledTaskSettingsSet,New-ScheduledTaskTrigger,New-ScriptFileInfo,New-SmbGlobalMapping,New-SmbMapping,New-SmbMultichannelConstraint,New-SmbShare,New-StorageFileServer,New-StoragePool,New-StorageSubsystemVirtualDisk,New-StorageTier,New-TemporaryFile,New-VirtualDisk,New-VirtualDiskClone,New-VirtualDiskSnapshot,New-Volume,New-VpnServerAddress,Open-NetGPO,Optimize-StoragePool,Optimize-Volume,oss,Pause,prompt,PSConsoleHostReadline,Publish-Module,Publish-Script,Read-PrinterNfcTag,Register-ClusteredScheduledTask,Register-DnsClient,Register-IscsiSession,Register-PSRepository,Register-ScheduledTask,Register-StorageSubsystem,Remove-AutologgerConfig,Remove-BitLockerKeyProtector,Remove-DAEntryPointTableItem,Remove-DnsClientNrptRule,Remove-DscConfigurationDocument,Remove-DtcClusterTMMapping,Remove-EtwTraceProvider,Remove-FileShare,Remove-InitiatorId,Remove-InitiatorIdFromMaskingSet,Remove-IscsiTargetPortal,Remove-MaskingSet,Remove-MpPreference,Remove-MpThreat,Remove-NetAdapterAdvancedProperty,Remove-NetEventNetworkAdapter,Remove-NetEventPacketCaptureProvider,Remove-NetEventProvider,Remove-NetEventSession,Remove-NetEventVFPProvider,Remove-NetEventVmNetworkAdapter,Remove-NetEventVmSwitch,Remove-NetEventVmSwitchProvider,Remove-NetEventWFPCaptureProvider,Remove-NetFirewallRule,Remove-NetIPAddress,Remove-NetIPHttpsCertBinding,Remove-NetIPHttpsConfiguration,Remove-NetIPseospSetting,Remove-NetIPsecMainModeCryptoSet,Remove-NetIPsecMainModeRule,Remove-NetIPsecMainModeSA,Remove-NetIPsecPhase1AuthSet,Remove-NetIPsecPhase2AuthSet,Remove-NetIPsecQuickModeCryptoSet,Remove-NetIPsecQuickModeSA,Remove-NetIPsecRule,Remove-NetLbfoTeam,Remove-NetLbfoTeamMember,Remove-NetLbfoTeamNic,Remove-NetNat,Remove-NetNatExternalAddress,Remove-NetNatStaticMapping,Remove-NetNatTransitionConfiguration,Remove-NetNeighbor,Remove-NetQosPolicy,Remove-NetRoute,Remove-NetSwitchTeam,Remove-NetSwitchTeamMember,Remove-NetTransportFilter,Remove-NetworkSwitchEthernetPortIPAddress,Remove-NetworkSwitchVlan,Remove-Odbsn,Remove-Partition,Remove-PartitionAccessPath,Remove-PhysicalDisk,Remove-Printer,Remove-PrinterDriver,Remove-PrinterPort,Remove-PrintJob,Remove-SmbBandwidthLimit,Remove-SmbGlobalMapping,Remove-SmbMapping,Remove-SmbMultichannelConstraint,Remove-SmbShare,Remove-StorageFaultDomain,Remove-StorageFileServer,Remove-StorageHealthIntent,Remove-StorageHealthSetting,Remove-StoragePool,Remove-StorageTier,Remove-TargetPortFromMaskingSet,Remove-VirtualDisk,Remove-VirtualDiskFromMaskingSet,Remove-VpnConnection,Remove-VpnConnectionRoute,Remove-VpnConnectionTriggerApplication,Remove-VpnConnectionTriggerDnsConfiguration,Remove-VpnConnectionTriggerTrustedNetwork,Rename-DAEntryPointTableItem,Rename-MaskingSet,Rename-NetAdapter,Rename-NetFirewallRule,Rename-NetIPHttpsConfiguration,Rename-NetIPsecMainModeCryptoSet,Rename-NetIPsecMainModeRule,Rename-NetIPsecPhase1AuthSet,Rename-NetIPsecPhase2AuthSet,Rename-NetIPsecQuickModeCryptoSet,Rename-NetIPsecRule,Rename-NetLbfoTeam,Rename-NetSwitchTeam,Rename-Printer,Repair-FileIntegrity,Repair-VirtualDisk,Repair-Volume,Reset-DAClientExperienceConfiguration,Reset-DAEntryPointTableItem,Reset-DtcLog,Reset-NCSIPolicyConfiguration,Reset-Net6to4Configuration,Reset-NetAdapterAdvancedProperty,Reset-NetDnsTransitionConfiguration,Reset-NetIPHttpsConfiguration,Reset-NetIsatapConfiguration,Reset-NetTeredoConfiguration,Reset-PhysicalDisk,Reset-StorageReliabilityCounter,Resize-Partition,Resize-StorageTier,Resize-VirtualDisk,Restart-NetAdapter,Restart-PcsvDevice,Restart-PrintJob,Restore-DscConfiguration,Restore-NetworkSwitchConfiguration,Resume-BitLocker,Resume-PrintJob,Revoke-FileShareAccess,Revoke-SmbShareAccess,SafeGetCommand,Save-EtwTraceSession,Save-Module,Save-NetGPO,Save-NetworkSwitchConfiguration,Save-Script,Send-EtwTraceSession,Set-AutologgerConfig,Set-ClusteredScheduledTask,Set-DAClientExperienceConfiguration,Set-DAEntryPointTableItem,Set-Disk,Set-DnsClient,Set-DnsClientGlobalSetting,Set-DnsClientNrptGlobal,Set-DnsClientNrptRule,Set-DnsClientServerAddress,Set-DtcAdvancedHostSetting,Set-DtcAdvancedSetting,Set-DtcClusterDefault,Set-DtcClusterTMMapping,Set-Dtefault,Set-DtcLog,Set-DtcNetworkSetting,Set-DtcTransaction,Set-DtcTransactionsTraceSession,Set-DtcTransactionsTraceSetting,Set-DynamicParameterVariables,Set-EtwTraceProvider,Set-FileIntegrity,Set-FileShare,Set-FileStorageTier,Set-InitiatorPort,Set-IscsiChapSecret,Set-LogProperties,Set-MMAgent,Set-MpPreference,Set-NCSIPolicyConfiguration,Set-Net6to4Configuration,Set-NetAdapter,Set-NetAdapterAdvancedProperty,Set-NetAdapterBinding,Set-NetAdapterChecksumOffload,Set-NetAdapterEncapsulatedPacketTaskOffload,Set-NetAdapterIPsecOffload,Set-NetAdapterLso,Set-NetAdapterPacketDirect,Set-NetAdapterPowerManagement,Set-NetAdapterQos,Set-NetAdapterRdma,Set-NetAdapterRsc,Set-NetAdapterRss,Set-NetAdapterSriov,Set-NetAdapterVmq,Set-NetConnectionProfile,Set-NetDnsTransitionConfiguration,Set-NetEventPacketCaptureProvider,Set-NetEventProvider,Set-NetEventSession,Set-NetEventVFPProvider,Set-NetEventVmSwitchProvider,Set-NetEventWFPCaptureProvider,Set-NetFirewallAddressFilter,Set-NetFirewallApplicationFilter,Set-NetFirewallInterfaceFilter,Set-NetFirewallInterfaceTypeFilter,Set-NetFirewallPortFilter,Set-NetFirewallProfile,Set-NetFirewallRule,Set-NetFirewallSecurityFilter,Set-NetFirewallServiceFilter,Set-NetFirewallSetting,Set-NetIPAddress,Set-NetIPHttpsConfiguration,Set-NetIPInterface,Set-NetIPseospSetting,Set-NetIPsecMainModeCryptoSet,Set-NetIPsecMainModeRule,Set-NetIPsecPhase1AuthSet,Set-NetIPsecPhase2AuthSet,Set-NetIPsecQuickModeCryptoSet,Set-NetIPsecRule,Set-NetIPv4Protocol,Set-NetIPv6Protocol,Set-NetIsatapConfiguration,Set-NetLbfoTeam,Set-NetLbfoTeamMember,Set-NetLbfoTeamNic,Set-NetNat,Set-NetNatGlobal,Set-NetNatTransitionConfiguration,Set-NetNeighbor,Set-NetOffloadGlobalSetting,Set-NetQosPolicy,Set-NetRoute,Set-NetTCPSetting,Set-NetTeredoConfiguration,Set-NetUDPSetting,Set-NetworkSwitchEthernetPortIPAddress,Set-NetworkSwitchPortMode,Set-NetworkSwitchPortProperty,Set-NetworkSwitchVlanProperty,Set-Odbriver,Set-Odbsn,Set-Partition,Set-PcsvDeviceBootConfiguration,Set-PcsvDeviceNetworkConfiguration,Set-PcsvDeviceUserPassword,Set-PhysicalDisk,Set-PrintConfiguration,Set-Printer,Set-PrinterProperty,Set-PSRepository,Set-ResiliencySetting,Set-ScheduledTask,Set-SmbBandwidthLimit,Set-SmbClientConfiguration,Set-SmbPathAcl,Set-SmbServerConfiguration,Set-SmbShare,Set-StorageFileServer,Set-StorageHealthSetting,Set-StoragePool,Set-StorageProvider,Set-StorageSetting,Set-StorageSubSystem,Set-StorageTier,Set-TestInconclusive,Setup,Set-VirtualDisk,Set-Volume,Set-VolumeScrubPolicy,Set-VpnConnection,Set-VpnConnectionIPsecConfiguration,Set-VpnConnectionProxy,Set-VpnConnectionTriggerDnsConfiguration,Set-VpnConnectionTriggerTrustedNetwork,Should,Show-NetFirewallRule,Show-NetIPsecRule,Show-VirtualDisk,Start-AppBackgroundTask,Start-AutologgerConfig,Start-Dtc,Start-DtcTransactionsTraceSession,Start-EtwTraceSession,Start-MpScan,Start-MpWDOScan,Start-NetEventSession,Start-PcsvDevice,Start-ScheduledTask,Start-StorageDiagnosticLog,Start-Trace,Start-WUScan,Stop-DscConfiguration,Stop-Dtc,Stop-DtcTransactionsTraceSession,Stop-EtwTraceSession,Stop-NetEventSession,Stop-PcsvDevice,Stop-ScheduledTask,Stop-StorageDiagnosticLog,Stop-StorageJob,Stop-Trace,Suspend-BitLocker,Suspend-PrintJob,Sync-NetIPsecRule,TabExpansion2,Test-Dtc,Test-NetConnection,Test-ScriptFileInfo,Unblock-FileShareAccess,Unblock-SmbShareAccess,Uninstall-Dtc,Uninstall-Module,Uninstall-Script,Unlock-BitLocker,Unregister-AppBackgroundTask,Unregister-ClusteredScheduledTask,Unregister-IscsiSession,Unregister-PSRepository,Unregister-ScheduledTask,Unregister-StorageSubsystem,Update-Disk,Update-DscConfiguration,Update-EtwTraceSession,Update-HostStorageCache,Update-IscsiTarget,Update-IscsiTargetPortal,Update-Module,Update-ModuleManifest,Update-MpSignature,Update-NetIPsecRule,Update-Script,Update-ScriptFileInfo,Update-SmbMultichannelConnection,Update-StorageFirmware,Update-StoragePool,Update-StorageProviderCache,Write-DtcTransactionsTraceSession,Write-PrinterNfcTag,Write-VolumeCache
    },
    morekeywords={Do,Else,For,ForEach,Function,If,In,Until,While},
    alsodigit={-},
    sensitive=false,
    morecomment=[l]{\#},
    morecomment=[n]{<\#}{\#>},
    morestring=[b]{"}, %
    morestring=[b]{'},
    morestring=[s]{@'}{'@},
    morestring=[s]{@"}{"@} %
}
 
\hypersetup{
    colorlinks=true,
    linkcolor=blue,
    filecolor=magenta,
    urlcolor=cyan,
    citecolor=blue
}

\renewcommand{\sectionautorefname}{Section}

\usepackage{xfp}        %
\usepackage{siunitx}    %
\newcommand{\rounding}[1]{%
  \num[round-mode=places, round-precision=2]{\fpeval{#1}}%
}

\makeatletter
\newcommand{\rqcounterautorefname}{RQ\@gobble}

\makeatother

\definecolor{roma1}{HTML}{7E1700} %
\definecolor{roma2}{HTML}{995215} %
\definecolor{roma3}{HTML}{B0802B} %
\definecolor{roma4}{HTML}{C8B455} %
\definecolor{roma5}{HTML}{D0E3A3} %
\definecolor{roma6}{HTML}{A4E5D3} %
\definecolor{roma7}{HTML}{5DC1D3} %
\definecolor{roma8}{HTML}{3292C2} %
\definecolor{roma9}{HTML}{2064AE} %
\definecolor{roma10}{HTML}{033198} %

\definecolor{light-gray}{gray}{0.95}

\newcommand{\breaktt}[1]{{\urlstyle{tt}\nolinkurl{#1}}}

\newcommand{\qt}[1]{``#1''} %

\newcommand{\footurl}[1]{\footnote{\url{#1}}}

\ifELSEVIER{}
\newcommand{\refavail}{\nameref{sec:availability}}
\else
\newcommand{\refavail}{\autoref{sec:availability}}
\fi

\ifELSEVIER{}
\newcommand{\mapnote}{\ Map lines delineate study areas and do not necessarily depict accepted national boundaries; this applies to all maps in this paper.}
\else
\newcommand{\mapnote}{}
\fi

\ifELSEVIER{}
\renewcommand{\rowcolors}[3]{}
\renewcommand{\rowcolor}[1]{}
\fi

\newcommand{\TITLE}{A Data-Driven Analysis of Infostealer Malware Victims}

\lstdefinestyle{codestyle}{
  basicstyle=\ttfamily\scriptsize,
    numbers=left,
    numberstyle=\tiny\color{roma3},
    numbersep=5pt,
    xleftmargin=2em,
    framexleftmargin=2em,
    stepnumber=1,
    showstringspaces=false,
    breaklines=true,
    frame=single,
    backgroundcolor=\color{roma5!30},
    keywordstyle=\color{roma9},      %
    stringstyle=\color{roma2},       %
    commentstyle=\color{roma8},      %
}
\lstdefinelanguage{json}{
    morestring=[b]",                 %
    literate=
     *{0}{{{\color{roma3}0}}}{1}
      {1}{{{\color{roma3}1}}}{1}
      {2}{{{\color{roma3}2}}}{1}
      {3}{{{\color{roma3}3}}}{1}
      {4}{{{\color{roma3}4}}}{1}
      {5}{{{\color{roma3}5}}}{1}
      {6}{{{\color{roma3}6}}}{1}
      {7}{{{\color{roma3}7}}}{1}
      {8}{{{\color{roma3}8}}}{1}
      {9}{{{\color{roma3}9}}}{1}
      {:}{{{\color{roma1}{:}}}}{1}
      {,}{{{\color{roma1}{,}}}}{1}
      {\{}{{{\color{roma1}{\{}}}}{1}
      {\}}{{{\color{roma1}{\}}}}}{1}
      {[}{{{\color{roma1}{[}}}}{1} %
      {]}{{{\color{roma1}{]}}}}{1} %
      {true}{{{\color{roma1}true}}}{4}
      {false}{{{\color{roma1}false}}}{5}
      {null}{{{\color{roma1}null}}}{4}
}
\renewcommand{\sectionautorefname}{Section}

\ifELSEVIER{}

\fi

\newacronym{CSAM}{CSAM}{child sexual abuse material}
\newacronym{CSA}{CSA}{child sexual abuse}

\newacronym{LEA}{LEA}{law enforcement agency}
\newacronym{UCOL}{UCOL}{Undercover Online Operatives}

\ifELSEVIER\else
\title{\TITLE{}} %
\fi

\begin{document}

\ifELSEVIER{}
\begin{frontmatter}

\title{\TITLE{}}

\author[tuni]{Arttu Paju\corref{cor1}}
\ead{arttu.paju@tuni.fi}
\author[tuni]{Juha Nurmi}
\author[csic]{David Arroyo}
\author[csic]{Sergio Chica Manjarrez}
\author[urv,athena]{Fran Casino}
\author[cih]{Mikko Niemel\"{a}}
\author[mq]{Juuso Itkonen}
\author[utas]{Joel Scanlan}
\author[piraeus,athena]{Constantinos Patsakis}
\author[tudelft]{Georgios Smaragdakis}

\affiliation[tuni]{organization={Tampere University}, city={Tampere}, country={Finland}}
\affiliation[csic]{organization={Consejo Superior de Investigaciones Cient\'{\i}ficas (CSIC)}, city={Madrid}, country={Spain}}
\affiliation[urv]{organization={Universitat Rovira i Virgili}, city={Tarragona}, country={Spain}}
\affiliation[athena]{organization={Athena Research Center}, city={Athens}, country={Greece}}
\affiliation[cih]{organization={Cyber Intelligence House}, city={Singapore}, country={Singapore}}
\affiliation[mq]{organization={Macquarie University}, city={Sydney}, country={Australia}}
\affiliation[utas]{organization={University of Tasmania}, city={Hobart}, country={Australia}}
\affiliation[piraeus]{organization={University of Piraeus}, city={Piraeus}, country={Greece}}
\affiliation[tudelft]{organization={Delft University of Technology}, city={Delft}, country={The Netherlands}}

\cortext[cor1]{Corresponding author.}

\begin{abstract}

Infostealer malware infects devices worldwide and harvests their most sensitive contents: credentials,
browser sessions, private keys, and access certificates.
Yet its impact on victims remains difficult to study without an ethical, legal, and curated research dataset.
To close this gap, we build a privacy-preserving pipeline that turns illicitly sourced infostealer logs into a
reproducible research artifact, minimizing sensitive data while preserving measurement utility, and use it to
construct a dataset of 170,298 victims from logs of multiple infostealer families.
Analyzing these victims, we find that the most compromised services mirror the world's most popular platforms,
with gaming and entertainment services strongly overrepresented.
Within the sample we identify compromised credentials for high-value organizations, including law-enforcement
domains, government and military services, and all eight Ivy League universities, as well as substantial
exposure of security-critical infrastructure and of financial, remote-access, and development platforms.
Victims also show widespread credential reuse and significant revictimization risk, overlapping with phishing
and ransomware victim populations.
We release the first anonymized victim-level infostealer dataset under controlled access to enable
ethical, privacy-preserving, and
reproducible research on information security and victim behavior.
 \end{abstract}


\end{frontmatter}
\fi

\ifELSEVIER\else
\maketitle
\fi

\glsresetall{}

\section{Introduction}\label{sec:intro}
Recent threat intelligence identifies credential abuse as the top initial access vector, and an analysis of infostealer
credential logs found that 30\,\% of the compromised systems were identified as enterprise-licensed devices~\citep{Verizon}.
CrowdStrike's latest Global Threat Report estimates that 82\,\% of modern attacks are executed without malware,
with adversaries instead using hands-on-keyboard techniques that blend in with legitimate activity and impede
detection; it also reports a 70\,\% drop in average breakout time between 2021 and 2025, attributed to attackers
gaining access to valid accounts~\citep{crowdstrike}.

This wealth of valid credentials can be partially attributed to the rise of \textit{infostealers}, a class of malware designed to
harvest sensitive information from compromised devices. In 2024, IBM reported a 266\,\% increase in infostealer deployments and a 71\,\% increase in
valid credential attacks~\citep{ibm2024}. Infostealer malware collects credentials and personal data, often by exploiting full browser sessions with cookies, rendering
strong password policies and multi-factor authentication ineffective~\citep{CACM-infostealers}. Modern infostealers compromise systems far beyond simple credential theft, exfiltrating
private VPN configurations, digital certificates, Secure Shell (SSH) and Pretty Good Privacy (PGP) private keys, authentication tokens, and cryptocurrency
wallet keys. This broader data theft enables attackers to bypass security controls, impersonate users, and access
internal networks or digital assets, sharply escalating the impact of a compromise. For instance, the 2024
National Public Data (NPD) breach exposed
sensitive personally identifiable information (PII) such as full names, Social Security numbers (SSNs), addresses, birth dates, and phone numbers
of over 272 million citizens of the US, UK, and Canada~\citep{NPD-Krebs}. The primary factor in the attack was an infostealer
malware infection~\citep{NPD-infostealer, CACM-infostealers}.

\begin{table*}[!htb]
  \centering
  \caption{Research questions, their scope, and how they are answered.}\label{tbl:rqs}
  \scriptsize
  \rowcolors{2}{gray!10}{white}
  \renewcommand*{\arraystretch}{0.75}
  \begin{tabular}{lp{2.28in}p{3.43in}p{0.47in}}
    \toprule
    \textbf{\#} & \textbf{Research question} & \textbf{Scope} & \textbf{Answered}\\
    \midrule

    RQ1 & How can large-scale analysis of illicitly sourced infostealer logs be conducted and shared in ways that minimize harm, protect privacy, and support reproducible research?
    & While data anonymization can significantly impact data utility, we extract many insights about the victims of infostealers
    by using a minimized and anonymized set, providing a step-by-step methodology that can be replicated to
    analyze similar datasets and share the data with the community for further research. & \autoref{sec:methods}, \autoref{sec:ethics}, \refavail{} \\
    RQ2 & What tooling, techniques, and infrastructure do modern infostealers use to harvest and exfiltrate credential-rich logs?
    & We discuss the methods used to infiltrate hosts and the tactics, techniques, and procedures used after the
    analysis of thousands of samples of specific infostealer families. Additionally, we illustrate their infrastructure and C2 servers.
    &\autoref{sec:tooling}\\
    RQ3 & Who are the victims in the anonymized logs, what do their online behavior and credential practices reveal,
    and are they susceptible to other digital scams? &
    We characterize victim representativeness, geography, operating systems, and the service categories of
    victims' stored credentials. We then explore username and password practices and investigate whether
    victims have fallen for phishing attacks and to what extent recent ransomware victim organizations have
    leaked credentials.
    & \autoref{sec:representativeness}, \autoref{sec:geography}, \autoref{sec:os}, \autoref{sec:victim-authentication},
    \autoref{sec:revictimization}\\
    RQ4 & How often do infostealer logs harvest credentials associated with security critical systems like government,
    military, law enforcement, and critical infrastructure operators? &
    We analyze the collected logs to identify logs associated with mission critical systems, focusing on organizations
    that, if compromised, may have detrimental and cascading effects.
    &\autoref{sec:special_domains}, \autoref{sec:ip_addresses}\\
    RQ5 & What are the illegal activities of infostealer victims that can be inferred from the leaked logs?
    & Beyond reporting victims' service preferences, we identify domains directly linked to various
    forms of illegal activity.
    & \autoref{sec:crime}\\
    RQ6 & How exposed are victims' high-value commercial services, such as financial and communication platforms?
    & We quantify victims' exposed credentials for financial services and for communication and privacy-oriented platforms.
    & \autoref{sec:financial}, \autoref{sec:comm_privac}\\
    \bottomrule
  \end{tabular}
\end{table*}

While various forms of credential-stealing malware have existed for years,
dozens of infostealer families have recently emerged, offered
under the Malware-as-a-Service (MaaS)
model~\citep{patsakis2024malware} through various platforms,
significantly lowering the technical barrier for cybercrime operators.
In this context, after purchasing malware control software, the
adversary typically convinces victims (i.e., usually through social
engineering) to download and execute malicious binaries. Prior research has shown
that some infostealer logs are available online, even
though some sources are difficult to locate and
access~\citep{DBLP:conf/IEEEares/NurmiNB23}.

This paper complements previous studies on password stealers~\citep{AvgetidisAVLBKM23} by focusing on victims' characteristics and behavior.
Industry reporting draws on proprietary data that cannot be independently reproduced~\citep{ibm2024, crowdstrike}, and academic work has
largely examined the malware itself, the stolen-data markets, and their economics rather than the victims behind the
logs~\citep{DBLP:conf/IEEEares/NurmiNB23, patsakis2024malware, DBLP:conf/sp/MarjanovH25}.
This research gap is consequential: the logs expose not just individuals but the organizations they belong to, from critical infrastructure to
government services, and defenders cannot prioritize a risk they cannot measure.

We conduct a large-scale, data-driven analysis of the victims of modern infostealer malware.
We collect the logs from archives advertised as RedLine, Raccoon, Vidar Stealer, LummaC2, MetaStealer, StealC,
Rhadamanthys and Atomic macOS Stealer, some of which also bundle logs of unknown origin, so each label describes
an archive rather than any individual record.
To process them, we develop a legally and ethically approved methodology that removes direct personally
identifiable information while retaining analytical value and reducing residual re-identification risk.
We then analyze the anonymized data for insights into the victims and their exposure, paying particular
attention to the behavioral-analysis and impersonation risks~\citep{DBLP:conf/sp/MarjanovH25} that persist
even in an anonymized dataset derived from infostealer logs.

We constructed our dataset, which we share with the research community, from 170,298 distinct victims worldwide,
compromised in 2024--2025 and collected in January~2025.
We study the representativeness of our sample and find that, while it overrepresents an online gaming population, it still spans a broad, global demographic.

Building on this dataset, we examine not only infection demographics and exposed credentials but also victim interests, organizational
exposure, revictimization trends, and evidence of illegal or deviant behavior.
We find that cybercriminals themselves are sometimes infected, which enables us to study their illicit activity, and that
infostealers reach mission-critical infrastructure and services to a notable extent.
We summarize our research questions in \autoref{tbl:rqs}.

Our contributions can be summarized as follows:\\
(i) a data-driven analysis of infostealer victims and their exposure, to our knowledge the first
at this scale to study victims through openly shared logs rather than the malware or the markets
that sell it;\\
(ii) a methodology for the legally and ethically compliant investigation of infostealers; and\\
(iii) an analysis framework and the first anonymized victim-level infostealer dataset released for
reproducible research.

\autoref{sec:background} and \autoref{sec:methods} cover background and methodology, and \autoref{sec:tooling}
characterizes infostealer tooling and infrastructure.
\autoref{sec:results} reports our findings on the victims
and their exposure, \autoref{sec:revictimization} examines their exposure to other forms of cybercrime, and
\autoref{sec:discussion} interprets the findings, states our limitations, and draws lessons for defense.
\section{Background and related work}\label{sec:background}

Infostealer malware extracts sensitive data from infected devices and sends it to attacker-controlled infrastructure.
Infostealers are often sold as MaaS, allowing even unskilled actors to run credential theft campaigns~\citep{patsakis2024malware}.
They target browser-stored data, including credentials, cookies, and payment card information,
but also extract VPN configurations, cryptocurrency wallets, and sometimes even gaming or messaging tokens.
For example, Rhadamanthys can extract data from browsers, VPN/email clients, password managers (e.g., KeePass),
and crypto wallets~\citep{zscaler}.
Similarly, Lumma has specific modules for exfiltrating device data~\citep{lumma_trellix}.

Once active on a victim's machine, infostealers aggregate stolen data into log files.
Initially, the preferred exfiltration method was an HTTP POST to hardcoded C2 servers. More recently, many families have moved to Telegram bots~\citep{bitsight_telegram}, cloud application
programming interfaces (APIs), and abused services like Pastebin or GitHub to evade detection, channels also
seen in the samples we analyze (\autoref{tbl:ttps}). The resulting logs, often tagged with a unique \qt{bot ID,} form a highly lucrative dataset for attackers.
Moreover, operators may manually filter high-value credentials, such as VPNs or banking logins, for resale on underground markets.

Campaigns distributing infostealers typically rely on affiliates (also called \qt{traffers}), who use phishing~\citep{DBLP:journals/compsec/KoutsokostasLAO22},
malvertising, Search Engine Optimization (SEO) poisoning, and trojanized installers to infect systems~\citep{AvgetidisAVLBKM23,patsakis2024malware}.
Social platforms like TikTok and YouTube are also exploited through so-called \qt{ClickFix} scams, which masquerade as beneficial videos offering quick fixes to common tech issues,
yet deliver malware~\citep{tiktok}.
Newer delivery tactics include fake CAPTCHAs, where a hijacked page asks visitors to prove they are human by
pasting a command that the page has already copied to their clipboard, which runs PowerShell to fetch the payload.
A related trick packages the malware as a malicious dynamic-link library (DLL) beside a legitimate signed
executable that side-loads it, so the bundle only appears benign~\citep{v4ensicsLumma}.
More targeted lures also appear, such as malware disguised as a fake \qt{checker}, a tool marketed to other
criminals that claims to test lists of stolen credentials against a service and report which accounts still work.
Reported cases pose as an OnlyFans checker (LummaC2) or a Netflix
checker (RedLine)~\citep{cybereason_lummastealer,bitdef}. There are also reports of infostealers
embedded in browser extensions and malicious mobile apps~\citep{cybernjgov}.
This diverse set of infection approaches has led to
massive infection counts. For instance, Microsoft observed that the LummaC2 alone had over 394,000
global Windows infections in a two-month span in early 2025~\citep{ms}. Notably, infostealers are not only
targeting home users: enterprise employees are also targeted, since
corporate credentials (for VPNs, Single Sign-On accounts, etc.) can yield much higher value~\citep{bleepRussianMarket}.

In the past, stolen credentials were traded mostly on hacking forums or dark web marketplaces in a \qt{piecemeal} fashion.
The notorious \emph{Genesis Market} (2018--2023) exemplified this model, selling \qt{browser fingerprints} and account credentials from malware logs~\citep{DBLP:conf/IEEEares/NurmiNB23}.
Genesis allowed buyers to purchase a victim's entire identity
package (login cookies, saved passwords, and OS details) to
bypass multifactor authentication and fraud detection by perfectly mimicking the victim's device.
With the rise of specialized log markets and Telegram channels, this trade has become far more streamlined.
Telegram now serves both as an advertising hub for infostealer services and a marketplace where,
for a subscription fee, buyers can download large volumes of fresh logs without purchasing each victim's data individually (see \autoref{fig:adv} in~\autoref{app:offerings}).
Its popularity stems from ease of use, anonymity, and the ability to broadcast instantly to large audiences.

\subsection{Comparison with prior work}\label{sec:prior_work_comparison}

Previous infostealer studies have focused primarily on specific ecosystems or narrower datasets. For example,
Nurmi et al.~\citep{DBLP:conf/IEEEares/NurmiNB23} characterized the Genesis Market and its trading model but did not
analyze victim-level credential data at scale.
Patsakis et al.~\citep{patsakis2024malware} surveyed the MaaS landscape and categorized infostealer families, offering a
technical taxonomy but lacking a large-scale empirical measurement of victim logs.

The closest prior work to our study is Avgetidis et al.~\citep{AvgetidisAVLBKM23}.
For 20 months (April 2019 to December 2020), using a dataset built from tracking pixels embedded in decoy credentials,
they followed 4,586 stealer operator devices and 5,295 command-and-control (C2) server
panels (web interfaces on C2 servers where logs are exfiltrated) across 10 malware families.
They reported potential infections inside the networks of the U.S. Social Security Administration, the U.S. House of Representatives, and the U.S. Senate.
Their focus was the attacker's side, the operator devices and panel infrastructure.
The present study focuses on the victims and uses the stolen logs to quantify exposure of credentials and other information.

Detection work aimed specifically at infostealers is sparse. Buescher et al.~\citep{buescher2011banksafe} detected
credential theft at runtime from inside the browser, and Bustos-Tabernero et al.~\citep{bustos2023infostealer}
classified infostealer variants with graph neural networks. Behavioral profiling is more commonly done for malware in
general, with ATT\&CK serving as the vocabulary for describing what a sample does when executed~\citep{alsada2025attack}.
We do not propose a detector; \autoref{sec:tooling} characterizes what seven families do once they run.

Industry reports from IBM~\citep{ibm2024}, CrowdStrike~\citep{crowdstrike}, and others provide threat intelligence
based on proprietary data, which cannot be independently verified or reproduced.
Marjanov and Hutchings~\citep{DBLP:conf/sp/MarjanovH25} examined the risks associated with behavioral analysis from infostealer logs
but focused on re-identification rather than broad victim characterization.
Our work differs from these studies in three key aspects:
(i) we provide a large-scale, multi-family measurement (170,298 victims) from a single time window,
enabling cross-family technical and infrastructural comparisons;
(ii) we release an anonymized dataset and analysis scripts for full reproducibility, unlike proprietary industry reports; and
(iii) we examine not only technical indicators but also victim behavior, institutional exposure, revictimization, and digital hygiene,
bridging the gap between threat intelligence and empirical security measurement.
The sensitivity of this data demands a rigorous methodology that preserves victim privacy, which we present in the next section.
\section{Methodology}\label{sec:methods}

\begin{figure*}[!htb]
  \centering
  \includegraphics[width=\textwidth]{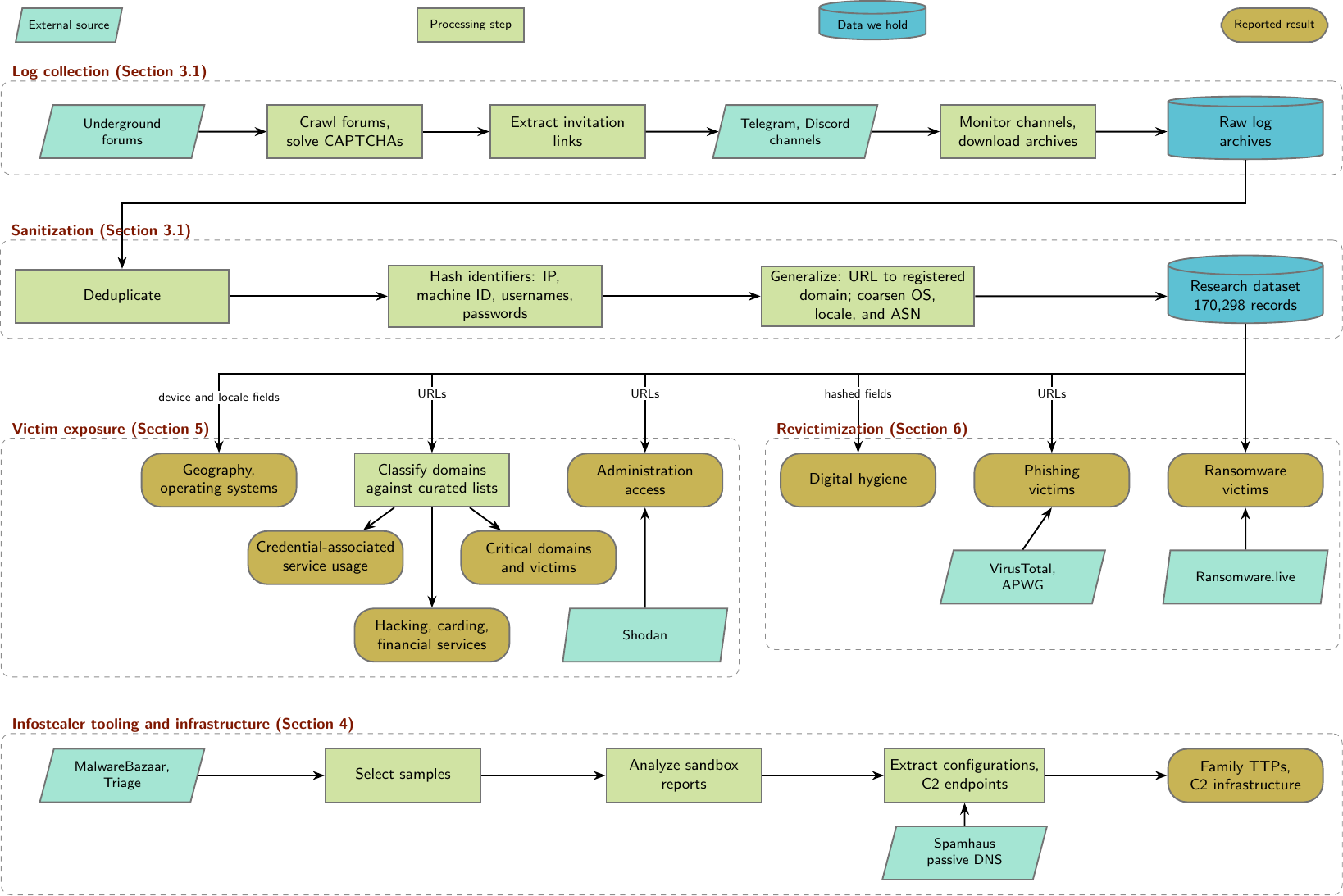}
  \caption{Pipeline of our methodology, from acquiring malware logs from illicit sources to the
  analyses each section reports.}\label{fig:methodology}
\end{figure*}

The scale of the infostealer ecosystem demands a privacy-first approach to studying victimization and organizational risk.
To address RQ1 and enable the rest of our analysis,
we designed the pipeline in \autoref{fig:methodology} to acquire, sanitize, and analyze infostealer
malware data exfiltrated from victim devices.
This pipeline received approval from the Ethics Advisory Board of the SafeHorizon project (see~\autoref{sec:ethics}).

\subsection{Data collection methodology}\label{sec:datacollection}

In 2024, we crawled 404 underground forums on Tor and the clear web.
These forums originated from an existing large-scale threat intelligence collection maintained by Cyber Intelligence House,
where analysts had verified each forum as having hosted, advertised, or repeatedly linked to infostealer
malware logs or related marketplaces.
We continuously crawled these forums to discover invitations to external communication channels,
which led us to monitor 53 Discord and Telegram channels that actively shared infostealer malware logs during January 2025.
These logs contained infostealer malware victim data from 2024 to 2025.
Each record carries a single date, the date the log was collected.
The source logs record neither when a stealer produced them nor when they were first published,
so the generation and publication periods cannot be recovered from our data.
42 of the monitored channels were hosted on Telegram, with the remaining 11 channels on Discord.
The channels covered multiple languages, primarily English, Russian, Ukrainian, and Chinese,
although additional languages were also observed.
Most Telegram channels were publicly accessible through invitation links,
while a smaller subset required private invitations.
\autoref{tbl:collected_logs} reports how the logs in our dataset are distributed over the archives we collected them
from. Each label is the family the archive was advertised as, not an attribution of the individual records it holds.
The dataset contains records from eight advertised families, but individual records carry no family
label; family-level analysis is therefore performed only on the malware samples
(\autoref{sec:tooling}), not on the victim records.

\begin{table}[!htb]
    \centering
    \caption{Collected logs per archive, labelled by the family the archive was advertised as.
    \qt{Collection} rows bundle multiple sources.}\label{tbl:collected_logs}
    \scriptsize
    \rowcolors{2}{gray!10}{white}
\begin{tabular}{p{1.5in}rrr}
\toprule
\textbf{Family} & \textbf{Logs} & \textbf{Usernames} & \textbf{Passwords}\\
\midrule
LummaC2 & 97,509 & 17,158,285 & 17,210,301 \\
Collection: Unknown, Atomic macOS Stealer, and Rhadamanthys & 59,336 & 9,203,402 & 9,525,152 \\
Vidar Stealer & 5,978 & 556,058 & 557,580 \\
StealC & 5,630 & 484,189 & 511,119 \\
Collection: Raccoon, LummaC2, MetaStealer, and StealC & 1,625 & 185,558 & 186,171 \\
RedLine & 168 & 24,190 & 24,207 \\
Collection logs: MetaStealer and StealC & 52 & 10,023 & 12,159 \\\midrule
\rowcolor{white}\textbf{Total} & 170,298 & 27,621,705 & 28,026,689 \\
\bottomrule
\end{tabular}
\end{table}

We collect data through a fully automated, iterative pipeline designed for reproducibility and minimal manual bias:

\textbf{1. Forum crawling:}
Dedicated crawlers continuously collect text from the manually curated underground forums on both the clear web and Tor to detect indicators of infostealer activity,
including malware promotion, advertisements for leaked logs, configuration leaks, and shared malware log archives.
These crawlers solve CAPTCHAs and authenticate themselves in forums to collect website text content.

\textbf{2. Telegram and Discord communities:}
Pattern-matching algorithms automatically identify invitation links to Telegram and Discord communities together with references to emerging channels connected to infostealer ecosystems.
Newly discovered channels are manually verified before being incorporated into the monitoring pipeline.

\textbf{3. Telegram and Discord monitoring:}
Automated bots join the verified data leak groups to maintain situational awareness and collect newly published data sources.
Some channels were dedicated almost exclusively to distributing infostealer malware logs, often on a daily or weekly basis,
whereas others shared malware logs only occasionally alongside broader discussions of leaked databases, credential dumps, and other cybercrime-related content.

\textbf{4. Infostealer malware log collection:}
Infostealer malware network operators share data for free (i.e., for advertisement purposes, see \autoref{fig:adv} in~\autoref{app:offerings}) both in messenger groups and on the underground forums.
When malware log archives or related files are detected, they are programmatically downloaded into a secure analysis environment.
These raw datasets are then subjected to a controlled preprocessing phase.

\textbf{5. Processing research datasets:}
Finally, we extracted and minimized the data to create a curated research dataset.
We removed direct personal identifiers while retaining the necessary behavioral and technical indicators for the analysis.
We replaced victims' IP addresses, usernames, passwords, and hardware identifiers by hashing each value with a long random salt.
We used the same salt for all logs during a single processing run, so identical input values produced identical
identifiers throughout the dataset, which allowed us to align records across logs.
For example, in the example entry in~\autoref{app:JSON}, the same usernames output the same value, \qt{b8f2935589.}
We did not retain the original values or the salt value; thus, the identifiers cannot be reversed to recover the original values.

The URL field recorded which services each victim had authenticated to, and its path, query, and fragment
carried account identifiers, session tokens, and the structure of private endpoints.
We therefore reduced every URL to its registered domain, keeping only the \breaktt{http://} or
\breaktt{https://} prefix and discarding any embedded credentials, the port, the path, the query, and the
fragment, so that \breaktt{https://login.example.com/user?token=abc} became \breaktt{https://example.com}.
Keeping the prefix preserves whether the victim reached the service over an encrypted connection.
Raw IP addresses and internal host names have no registered domain, so we kept the host itself for those.

After removing the direct identifiers, we generalized the data further, because a combination of operating
system, language, keyboard layouts, VPN status, and autonomous system could identify a victim even though
none of these fields identify anyone on its own.
We treated these five fields as quasi-identifiers and generalized them in two steps, leaving VPN status
unchanged throughout, because an aggregate label would have stated something untrue about the victim.
First, we replaced every value that appeared in fewer than 100 records with an aggregate label.
Second, we enforced $k$-anonymity with $k = 5$: we counted the occurrences of every combination of the five
fields, and for each combination occurring in fewer than five records we replaced further values with the
aggregate label until at least five records shared the combination.
We replaced the values in a fixed order, starting with the autonomous system, then the keyboard layouts, the
operating system, and finally the language.
The released dataset contains 3,470 distinct combinations of the five fields, each shared by at least five
records.

\autoref{tbl:field_treatment} lists the content, the disclosure risk, and the treatment of every field in a
released record.
We did not generalize country information, because it is the unit of analysis for the geographic results
in~\autoref{sec:results}.
Including it among the five fields above would have left \rounding{9023/170298*100}\,\% of records in
combinations occurring fewer than five times, a residual re-identification risk that the controlled access
described in~\autoref{sec:ethics} addresses.

\begin{table*}[!htb]
\centering
\caption{Field-level treatment of the released victim records. Direct identifiers were replaced by salted
hashes, URLs were reduced to their registered domain, and the quasi-identifiers were generalized to prevent
re-identification.}\label{tbl:field_treatment}
\scriptsize
\rowcolors{2}{gray!10}{white}
\begin{tabular}{p{1.05in}p{1.15in}p{1.55in}p{2.2in}}
\toprule
\textbf{Field} & \textbf{Content} & \textbf{Disclosure risk} & \textbf{Treatment in the released dataset}\\
\midrule
\breaktt{ip_address} & Victim IP address & Direct identifier & Salted hash \\
\breaktt{machineid} & Hardware identifier & Direct identifier & Salted hash \\
\breaktt{usernames}, \breaktt{passwords} & Credentials for the compromised services & Direct identifiers;
usernames frequently contain real names & Salted hash; no recoverable credential material remains \\
\breaktt{urls} & Services the victim authenticated to & Paths and query strings carry account identifiers,
session tokens, and the internal structure of private endpoints & Reduced to
\breaktt{http(s)://registered-domain}; credentials, port, path, query, and fragment are dropped \\
\breaktt{onion_domain_names} & Tor services the victim reached & As above & Host names only, as collected \\
\breaktt{ip_asn} & Autonomous system and its operator & A small network can identify an employer, a campus,
or a single subscriber & Kept for the 275 systems seen in at least 100 records, which covers
\rounding{108913/170298*100}\,\% of victims; the rest read \qt{ASN: aggregated} \\
\breaktt{os} & Operating system & The build number narrows the population sharply & Edition kept, build
number and architecture removed \\
\breaktt{language} & Locale of the compromised host & The region subtag is often rare & Reduced to the
language subtag, so \qt{es-UY} becomes \qt{es} \\
\breaktt{keyboard_layouts} & Keyboard layouts installed on the host & Multilingual layout sets are close to
unique & Reduced to layout language names, so \qt{English (India)} becomes \qt{English} \\
\breaktt{location} & City and region derived from the IP & Near unique in combination with the fields above &
Removed \\
\breaktt{is_vpn} & Whether the IP belongs to a VPN provider & Two values, low on its own & Kept \\
\breaktt{country}, \breaktt{ip_location} & ISO-2 country code & Coarse on its own & Kept; the unit of
analysis for the geographic results \\
\breaktt{timestamp} & Date the log was collected & Narrows the infection window & Kept; the source logs carry
no time of day \\
\breaktt{*_count} & Number of URLs, usernames, and passwords & Aggregate & Kept, and unchanged by the
treatments above \\
\bottomrule
\end{tabular}
\end{table*}

This structured approach supports compliance with the legal and ethical constraints on using illicitly sourced
data~\citep{DBLP:conf/imc/ThomasPHCB17, kohno};
\autoref{sec:ethics} discusses how we balance the benefits and risks for victims, organizations, security practitioners, and society.
Although we analyze only anonymized data and not the full logs, the pipeline (\autoref{fig:methodology}) enables us to curate and
study the dataset more systematically than current industry reports.

\autoref{fig:methodology} traces this pipeline end to end.
It begins with the underground-forum and Telegram and Discord log collection described above.
The resulting logs are deduplicated, minimized, and anonymized into the research dataset, which the
victim analyses of \autoref{sec:results} and \autoref{sec:revictimization} then study.
We also characterize the malware behind these logs (\autoref{sec:tooling}), drawing on the sample
records and sandbox reports published by \texttt{MalwareBazaar} and \texttt{Triage}, and
extracting their configurations and C2 infrastructure.

\subsection{Dataset representativeness and bias characterization}\label{sec:datacleaning} %

Our dataset is derived from infostealer logs and does not constitute a random sample of the global internet population.
Instead, it reflects the victims whose devices were compromised by specific infostealer campaigns and whose logs were subsequently shared online.
Within each victim's logs we observe only the services the victim authenticated to with stored credentials, so we do
not see the full browsing history and miss any site visited without logging in (\autoref{sec:limitations}).
To characterize the dataset, we compare the prevalence of widely used services in our dataset with global usage statistics and compute the infection rate per country.
The dataset is skewed toward a highly active online gaming population, pointing to the infection method for infostealer distribution (e.g., via cheats, mods, or pirated software).

We show that the dataset provides broad geographic coverage and captures a security-relevant online gamer subpopulation.
It is therefore suitable for studying credential compromise, victim behavior, and organizational exposure within this segment.
Future work can reduce bias by integrating additional data sources (e.g., incident response datasets, enterprise telemetry, or multiple log-sharing ecosystems).

After removing duplicates and victims without compromised services, we study a sample of distinct victims (N = 170,298).
These victims possess unique machine identifiers or distinct IP addresses and at least one compromised service.
The macOS victims lack machine identifiers; consequently, we compare their IP addresses.
Each of these victims is one record in the released dataset, and throughout this article one record is one victim: every
count we report is a count of records rather than of devices or of people.
The mapping is not exact, since a device on a dynamic address can appear as more than one record, while a shared address,
a VPN exit point or a public machine can merge several devices into one.
We bound this effect for the macOS victims in~\autoref{sec:limitations}.
\section{Infostealer tooling and infrastructure}\label{sec:tooling}

Before turning to the victims, we characterize the malware behind these logs, working from samples
rather than from the logs themselves: how seven of the eight families named
in~\autoref{tbl:collected_logs} operate and the infrastructure that supports them. Samples for
Atomic macOS Stealer were not available.
Once launched, an infostealer harvests sensitive data from installed browsers (passwords, cookies, and session
tokens), cryptocurrency wallets, communication platforms, gaming credentials, payment methods, local accounts, and
email, and it may log keystrokes, capture screenshots, and read the clipboard. These capabilities let attackers
bypass authentication and exfiltrate highly sensitive data, causing serious business, financial, or personal harm.

\textbf{Method:}
To characterize the infostealer families and address RQ2, we studied the behavior of infostealer
samples covering seven of these families, drawing on the records and sandbox reports published by
MalwareBazaar~\citep{bazaar} and Triage~\citep{triage}. The executions, the family labels, and the
mapping from observed behavior to MITRE ATT\&CK techniques~\citep{mitreattack} are those services'
own. %
The sample and variant counts in \autoref{tbl:ttps} cover 389,562 samples.
The technique rankings in its last column come from an earlier extraction of 59,451 of these
samples, and we therefore report the ten techniques each family triggers most often as an ordering
rather than as counts.
We then identified the families' command and control (C2)
servers and mapped where their infrastructure is hosted, extracting IP addresses where they were directly used and
resolving historical addresses through Spamhaus' passive DNS API~\citep{spamhaus}. We excluded abused benign services
such as \texttt{pastebin.com}, \texttt{telegram.org}, \texttt{steamcommunity.com}, and \texttt{raw.githubusercontent.com}.

\textbf{Results:}
The tactics, techniques, and procedures (TTPs) that dominate across all families are system-oriented:
T1614.001 (System Language Discovery), T1012 (Query Registry), T1082 (System Information Discovery), and
T1005 (Data from Local System) rank among the most common, reflecting infostealers' reliance on system data
access and discovery.

The families differ most in what they do around collection. Persistence ranks in the top ten for RedLine, Raccoon and
StealC, through registry run keys (T1060, T1547.001), scheduled tasks (T1053) and service creation (T1543.003), and
RedLine alone ranks disabling security tools (T1089). LummaC2, Vidar and Rhadamanthys stay within discovery and
collection, consistent with a single run that exfiltrates and exits. Virtualization and sandbox checks (T1497) rank
high only for LummaC2, StealC and Rhadamanthys, all first seen in 2022 or 2023. Two families stand apart:
Rhadamanthys is the only one to rank email collection (T1114), matching its targeting of mail and VPN
clients~\citep{zscaler}, and MetaStealer's ranks are dominated by installer and script execution (T1546.016,
T1218.007, T1059.001) together with permission changes (T1222), reflecting delivery through trojanized installer
packages. The mapping mixes current and superseded ATT\&CK identifiers, so Raccoon ranks both T1130 and T1553.004,
and RedLine both T1060 and T1547.001, for what is the same behavior.

\begin{figure}[!htb]
    \centering
    \includegraphics[width=\columnwidth]{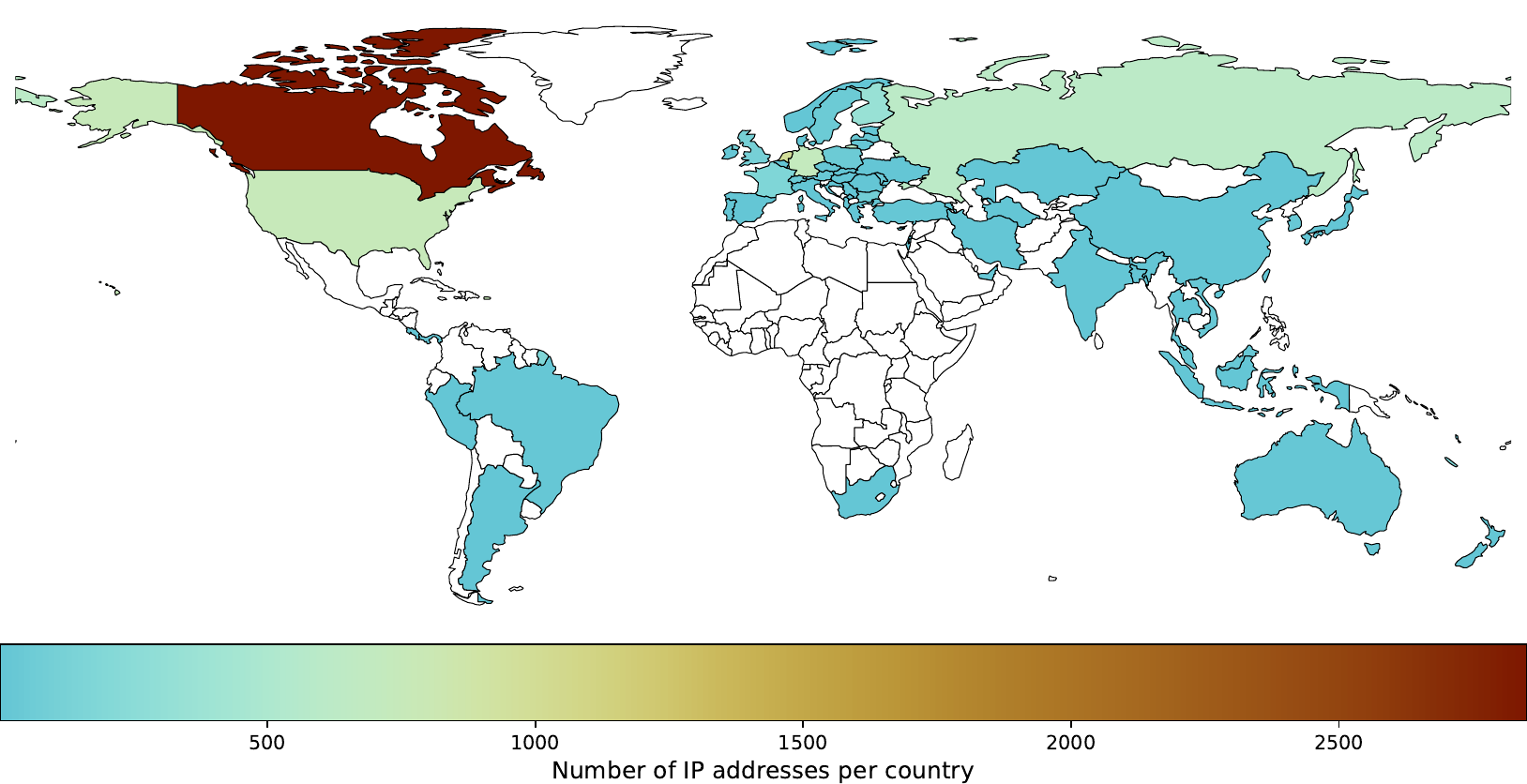}
    \caption{%
Infostealer C2 Infrastructure (partial).\mapnote{}}\label{fig:infra}
\end{figure}

For C2, some families used domain generation
algorithms~\citep{PlohmannYKBG16,CasinoLHPH21} or private Telegram channels, and some share common infrastructure
(\autoref{fig:commoninfra} in~\autoref{app:c2common}).
MetaStealer's algorithm, for instance, draws from a pool of about 60,000 \texttt{.xyz} domains.
Passive DNS shows which families resolved to an address, not
who leased it, so this is overlap in hosting rather than evidence about operators. In total, the hosting infrastructure spans 7,401 unique IP addresses (125 IPv6;
\autoref{fig:infra}), reflecting the industrialized scale of MaaS operations.

\begin{table*}[!htb]
    \centering
    \caption{Platforms, discovery date, and the ten most common TTPs of the seven infostealer families we
    obtained samples for, in
    decreasing order of frequency. The sample and variant counts cover 389,562 samples; the TTP
    rankings were computed over an earlier extraction of 59,451 of them.}\label{tbl:ttps}
    \scriptsize
    \rowcolors{2}{gray!10}{white}
    \resizebox{\textwidth}{!}{
    \begin{tabular}{rrrrrrp{0.3\linewidth}p{0.3\linewidth}}
        \toprule
        \textbf{Family} & \textbf{Platforms} & \textbf{First Seen} & \textbf{Variants} & \textbf{Samples} &\textbf{\#
Observed C2s} &\textbf{C2 types} & \textbf{TTPs} (MITRE ATT\&CK categorization~\citep{mitreattack})\\
        \midrule
        LummaC2 & Windows/macOS & August 2022 & 8,636 & 34,971 & 6,180 & Abuse of benign domains and services: \texttt{pastebin.com} 77 \texttt{.xyz} domains,
        1,224 \texttt{.shop} domains, 1,203 \texttt{.top}, 128 \texttt{.icu} & T1614.001, T1012, T1082, T1497, T1112, T1005,
        T1552.001, T1553.004, T1217, T1555.003\\
        MetaStealer & Windows/macOS & March 2023 & 6 & 103 & n/a & DGA candidate space of 60,000 \texttt{.xyz} domains, not observed endpoints & T1082, T1614.001, T1012, T1120,
        T1222, T1546.016, T1218.007, T1059.001, T1005, T1552.001\\
        Raccoon & Windows & April 2019 & 1,073 & 6,362 & 579 & 7 unique domains, no benign. 4 \texttt{.com}, 1 \texttt{.ink}, 1 \texttt{.xyz}, 1~.site
        & T1112, T1012, T1005, T1081, T1130, T1082, T1553.004, T1120, T1060, T1053\\
        RedLine & Windows & March 2020 & 7,732 & 287,103 & 3,293 & Abuse of benign domains and services: \texttt{duckdns.org}, \texttt{ddns.net}, \texttt{pastebin.com},
        \texttt{ngrok.io}, Telegram (\texttt{t.me}), 163 \texttt{.xyz} domains, 69 \texttt{.top} domains, 11 \texttt{.shop} domains & T1005, T1012,
        T1081, T1082, T1112, T1060, T1053, T1031, T1089, T1547.001\\
        Rhadamanthys & Windows & December 2022 & 259 & 1,432 & 158 & 15 unique domains, no benign. 1 \texttt{.asia}, 3 \texttt{.xyz}, 1 \texttt{.nl}, 1~.site,
        2 \texttt{.online}, 1 \texttt{.shop}, 2 \texttt{.ug}, 3 \texttt{.com}, 1 \texttt{.ug}, 1 \texttt{.net} & T1082, T1012, T1614.001, T1114,
        T1120, T1497, T1005, T1112, T1081, T1059.001\\
        StealC & Windows & January 2023 & 968 & 32,864 & 553 & Abuse of benign domains and services: Telegram (\texttt{t.me}) and \texttt{steamcommunity.com} 3
        \texttt{.xyz} domains, 28 \texttt{.top} domains & T1082, T1012, T1614.001, T1497, T1005, T1552.001, T1112, T1120, T1547.001, T1543.003\\
        Vidar Stealer & Windows & October 2018 & 2,244 & 26,727 & 272 & Abuse of benign domains and services: (\texttt{tumblr.com}), GitHub (\texttt{github.com} and \texttt{githubusercontent.com}),
        \texttt{steamcommunity.com}, \texttt{tiktok.com}, \texttt{ultimate-guitar.com}, Telegram (\texttt{t.me}), \texttt{mastodon.social}, 15 \texttt{.xyz} domains & T1082, T1012,
        T1112, T1005, T1614.001, T1552.001, T1553.004, T1217, T1555.003, T1081\\
        \bottomrule
    \end{tabular}}
\end{table*}

\section{Analysis of infostealer victims and their exposure}\label{sec:results}

The methodology of \autoref{sec:methods} produced an anonymized dataset we can study without
exposing victims. We use it first to characterize the victims themselves, and then to measure what
their compromise exposes: the organizations whose credentials they hold, their own illegal
activity, and their high-value personal services.

\subsection{Representativeness}\label{sec:representativeness}

Our logs cover only the people who were infected, not internet users at large, so any conclusion we draw rests on
knowing who those victims are. Characterizing them is the first half of RQ3. The dataset reaches victims in all
major countries (\autoref{sec:geography}); here we begin by asking whether the services they use match the wider
online population.

\textbf{Method:}
We compare how often popular domains appear among victims against publicly
available global usage statistics.

\textbf{Results:}
The most frequently compromised services prove to be the globally most-used online
services. We compute the main domains from the compromised services and the share of victims holding credentials for
each (see \autoref{tbl:domains}). Every one of these top 40 domains is held by more than 5\,\% of victims,
including roughly half who hold Facebook (Meta) and Google credentials (Gmail users authenticate to Google).

\begin{table}[!htb]
  \centering
  \footnotesize
  \caption{The top 40 domains from the compromised services and the percentage of victims with credentials to these services.}\label{tbl:domains}
  \rowcolors{2}{gray!10}{white}
  \begin{tabular}{p{2.5cm}rp{2.5cm}r}
    \toprule
    \textbf{Domain} & \textbf{\,\%} & \textbf{Domain} & \textbf{\,\%} \\
    \midrule
google.com & 64.04 & github.com & 10.14\\
facebook.com & 51.30 & openai.com & 9.77\\
live.com & 46.93 & yahoo.com & 9.30\\
discord.com & 30.02 & zoom.us & 8.99\\
instagram.com & 29.46 & rockstargames.com & 8.44\\
roblox.com & 28.61 & tlauncher.org & 8.10\\
netflix.com & 25.96 & sony.com & 7.38\\
amazon.com & 20.49 & adobe.com & 6.76\\
steampowered.com & 19.39 & ubisoft.com & 6.48\\
twitter.com x.com & 18.12 & nexusmods.com & 6.17\\
paypal.com & 16.96 & localhost & 6.13\\
spotify.com & 16.20 & tiktok.com & 6.10\\
microsoftonline.com & 16.17 & samsung.com & 5.93\\
twitch.tv & 15.99 & battle.net & 5.57\\
apple.com & 15.98 & ea.com & 5.46\\
epicgames.com & 14.56 & xiaomi.com & 5.39\\
riotgames.com & 13.97 & snapchat.com & 5.36\\
mega.nz & 12.57 & aliexpress.com & 5.18\\
steamcommunity.com & 11.97 & disneyplus.com & 5.16\\
linkedin.com & 11.38 & aternos.org & 5.11\\
     \bottomrule
  \end{tabular}
\end{table}

\autoref{tab:usage_comparison} shows each domain's prevalence among victims against its global usage: the share of victims
holding credentials, the estimated global usage of the service (as a percentage and in millions of monthly active users),
and the relative difference between the two.

\begin{table}[!htb]
\centering \footnotesize
\caption{Comparison of domain prevalence in infostealer data (\%)
  vs.\ global usage (\% and millions), and the infostealer victim data
  relative difference (\%).}\label{tab:usage_comparison}
\rowcolors{2}{gray!10}{white}
\begin{tabular}{lccr}
\toprule \textbf{Domain} & \textbf{Victims} & \textbf{Global usage} &
\textbf{Difference} \\ \midrule
google.com (gmail)   & 64.04\,\% & 44.96\,\% (2,500~\citep{gmail_users}) & +42\,\% \\
facebook.com         & 51.30\,\% & 55.13\,\% (3,065~\citep{backlinko_facebook_users}) & -7\,\% \\
instagram.com        & 29.46\,\% & 35.97\,\% (2,000~\citep{backlinko_instagram_users}) & -18\,\% \\
discord.com          & 30.02\,\% & 3.60\,\% (200~\citep{businessofapps_discord_statistics}) & +734\,\% \\
roblox.com           & 28.61\,\% & 6.83\,\% (380~\citep{backlinko_roblox_users}) & +319\,\% \\
netflix.com          & 25.96\,\% & 5.43\,\% (302~\citep{demandsage_netflix_subscribers}) & +378\,\% \\
amazon.com           & 20.49\,\% & 5.58\,\% (310~\citep{yaguara_amazon_statistics}) & +267\,\% \\
steampowered.com     & 19.39\,\% & 2.37\,\% (132~\citep{demandsage_steam_statistics}) & +718\,\% \\
twitter.com x.com    & 18.12\,\% & 11.98\,\% (666~\citep{backlinko_twitter_users}) & +51\,\% \\
paypal.com           & 16.96\,\% & 7.81\,\% (434~\citep{chargeflow_paypal_statistics}) & +117\,\% \\
spotify.com          & 16.20\,\% & 12.14\,\% (675~\citep{backlinko_spotify_users}) & +33\,\% \\
\bottomrule
\end{tabular}
\end{table}

The victim sample and the global usage metrics do not represent the
same sample space, so \autoref{tab:usage_comparison} is descriptive
context rather than a statistical comparison. The relative differences
(such as +734\,\% for Discord or +718\,\% for Steam) indicate empirical
overrepresentation within the population affected by infostealers,
rather than a formal deviation from a homogeneous probabilistic
distribution, and reflect exposure to risk associated with specific
digital behaviors such as gaming and youth communication platforms.
The higher number of accounts for these services among victims
indicates that infostealers might spread through the typical behaviors
of gaming communities, which prior work has
analyzed~\citep{graw07,vakilinia26}. URLs containing the keyword
\qt{cheat} reach more than 8,500 victims with credentials for video
game cheat sites such as \textit{lavicheats} or \textit{skycheats}.
Plausible vectors include pirated game mods, cheat software,
unofficial game launchers, and social engineering on gamer-focused
platforms like Discord.

\begin{takeout}
  The most compromised services mirror the world's most popular platforms, yet gaming and
  entertainment services (Discord, Roblox, Netflix, Steam) are massively overrepresented, possibly
  pointing to game-related infection vectors such as cheats and pirated mods.
\end{takeout}

\subsection{Geographic distribution}\label{sec:geography}

\textbf{Method:}
We compare the number of infections to the number of internet users in each country.
As of early 2025, approximately 5.56 billion people worldwide are internet users, accounting for 68.7\,\% of the global population~\citep{datareportal2025}.

\begin{figure}[!htb]
    \includegraphics[width=1.0\linewidth]{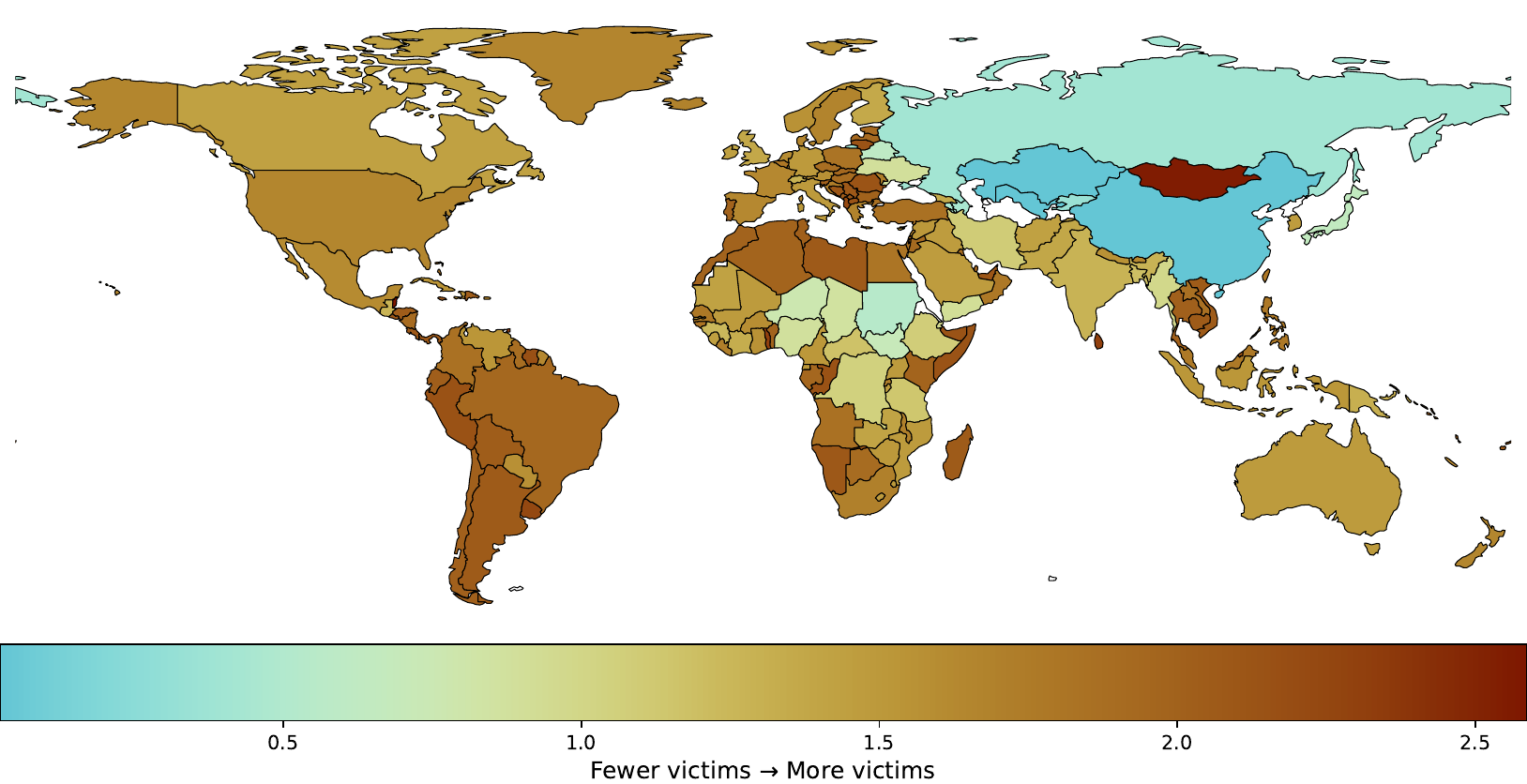}
    \caption{
    The number of victims per country per million internet users is calculated using a base-10 logarithmic scale.}\label{fig:victims_map}
\end{figure}

\textbf{Results:}
Of the 170,298 victims, 164,163 carry country information derived from their IP addresses,
and we observe victims from all major countries (see \autoref{fig:victims_map}).
Normalizing by internet population, the highest per-capita infection rates concentrate across Asia, Africa, Latin America, and Eastern Europe.
Among countries with at least 100 victims, Mongolia has the highest rate (260 infections, \rounding{356.5375269460092} per million internet users), followed by
Trinidad and Tobago (230 infections, \rounding{217.2385392502815} per million), and Togo (198 infections, \rounding{205.4378783870014} per million).
At the other extreme, Russia's 335 infections equate to only \rounding{2.5808936825885977} per million.
Eritrea, Tajikistan, Turkmenistan, and North Korea show no infections.

\begin{figure}[!htb]
  \includegraphics[width=\linewidth]{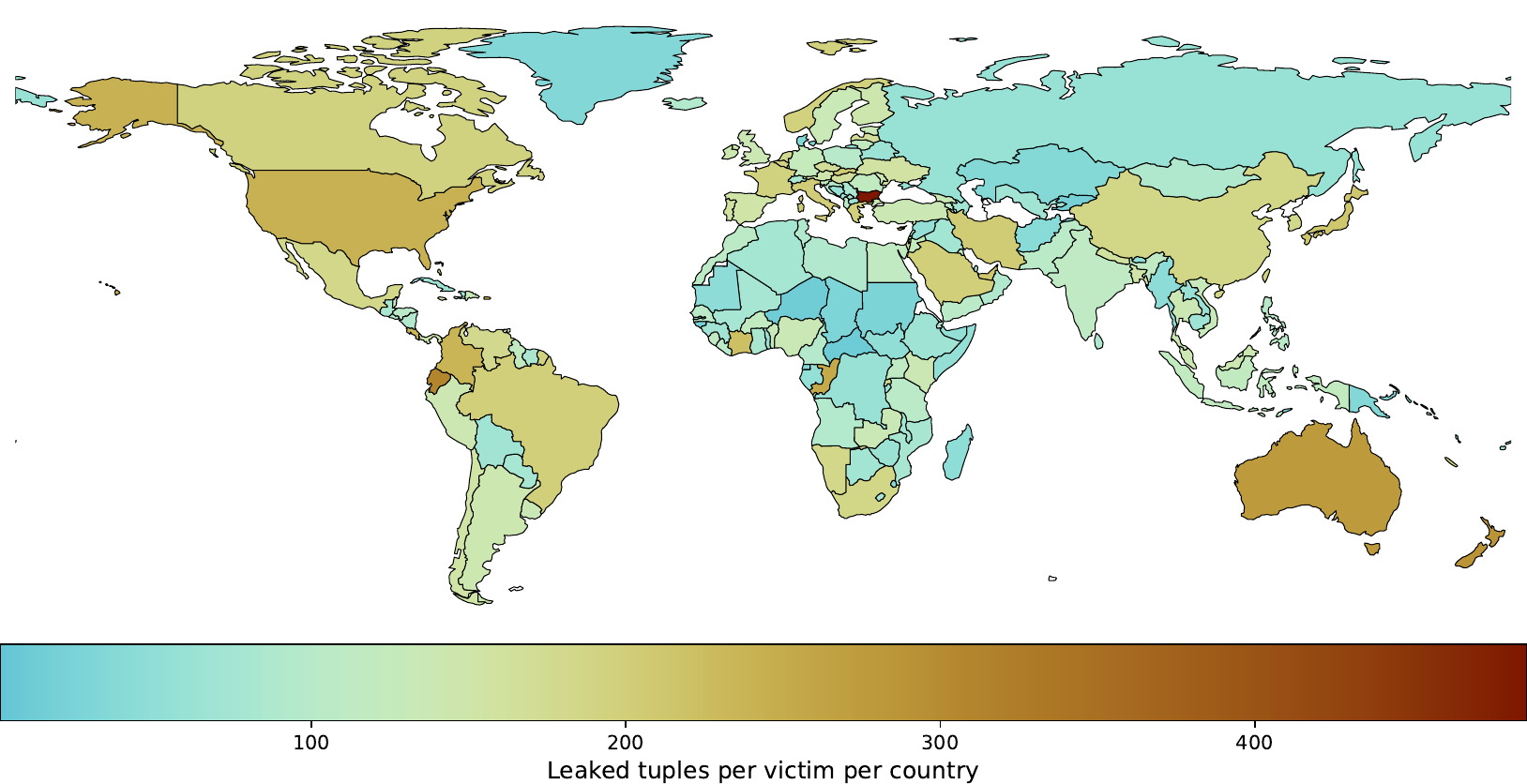}
  \caption{Average leaks per victim per country.}\label{fig:avg_disclosed_logs}
\end{figure}

The volume of leaked credentials also varies sharply across countries.
Each victim discloses, on average, 147.09 tuples of URL, username, and password, but
\autoref{fig:avg_disclosed_logs} shows large differences between countries.
These differences largely follow the reach of individual campaigns. Russian cybercriminals, for example,
run a large-scale ecosystem~\citep{russian_eco} built on infostealers such as Lumma; because these tools spread
mainly through Russian-speaking forums and are run by regional developers~\citep{lumma,lumma2}, their campaigns
stay geographically concentrated, which in turn shapes who their victims are.
A few hosts leak far more credentials than the rest: 70 hosts exceed 10,000 tuples, two orders of magnitude above the average,
and the largest single leak holds 157,107 tuples. These outliers are most likely shared machines in public
places such as libraries or airports, where many people save their credentials on the same device.

\begin{takeout}
  Per-capita infection rates vary widely across countries and are highest across Asia, Africa, Latin America, and Eastern Europe.
  The number of credentials leaked per victim follows the reach of individual campaigns.
\end{takeout}

\subsection{Operating systems}\label{sec:os}

\textbf{Method:}
We tabulate the operating system reported in each victim's log, standardizing names across build numbers and
language variants and omitting architecture, since almost all systems are 64-bit.

\textbf{Results:}
\autoref{tbl:victim_os} shows the operating systems the victims run. Windows dominates, and only a minimal fraction
of victims (\(\approx\) 1.37\,\%) run deprecated systems such as Windows 7 and 8 that no longer receive official
security updates~\citep{eol2024}; macOS accounts for 4,523 victims (2.66\,\%).
The logs record the operating system release, not its patch level, so these counts show which releases
victims run but not how current those installations were kept.
\autoref{sec:family_strata} compares the victims of the two platforms.

\begin{table}[!tbp]
  \centering
  \scriptsize
  \caption{Operating systems used by the victims.}\label{tbl:victim_os}
\rowcolors{2}{gray!10}{white}
  \begin{tabular}{lrrlrr}
    \toprule
    \textbf{Operating system} & \textbf{Count} & \textbf{\%} & \textbf{Operating system} & \textbf{Count} &\textbf{\%}\\
    \midrule
Win 10 Pro & 65,519 & 38.47 & Windows 11 & 4,848 & 2.85\\
Win 11 Home & 27,096 & 15.91 & Win 10 Enterprise & 4,807 & 2.82\\
Win 11 Pro & 22,606 & 13.27 & macOS & 4,523 & 2.66\\
Win 10 Home & 20,735 & 12.18 & Windows 10 & 1,755 & 1.03\\
Unknown & 10,122 & 5.94 & Win 7 Ultimate & 1,041 & 0.61\\
Other & 6,281 & 3.69 & Windows 8 & 965 & 0.57\\
\bottomrule
  \end{tabular}
\end{table}

\begin{takeout}
  Victims overwhelmingly run current, supported operating system releases, so no platform is safe from
  infection. Whether unpatched systems are over-represented is a separate question our data cannot answer.
\end{takeout}

\subsection{Victims' credential-associated service usage}\label{sec:victim-authentication}

We identify 322 victims authenticated to more than 500 unique domains (hosting one or more
services), a sign of infrequent password clean-up, a very long browsing history, or other practices
that leave them vulnerable to credential theft. This section asks what types of service the
credentials stolen from these victims belong to.

\begin{figure}[!htb]
            \centering
        \includegraphics[width=1.0\linewidth]{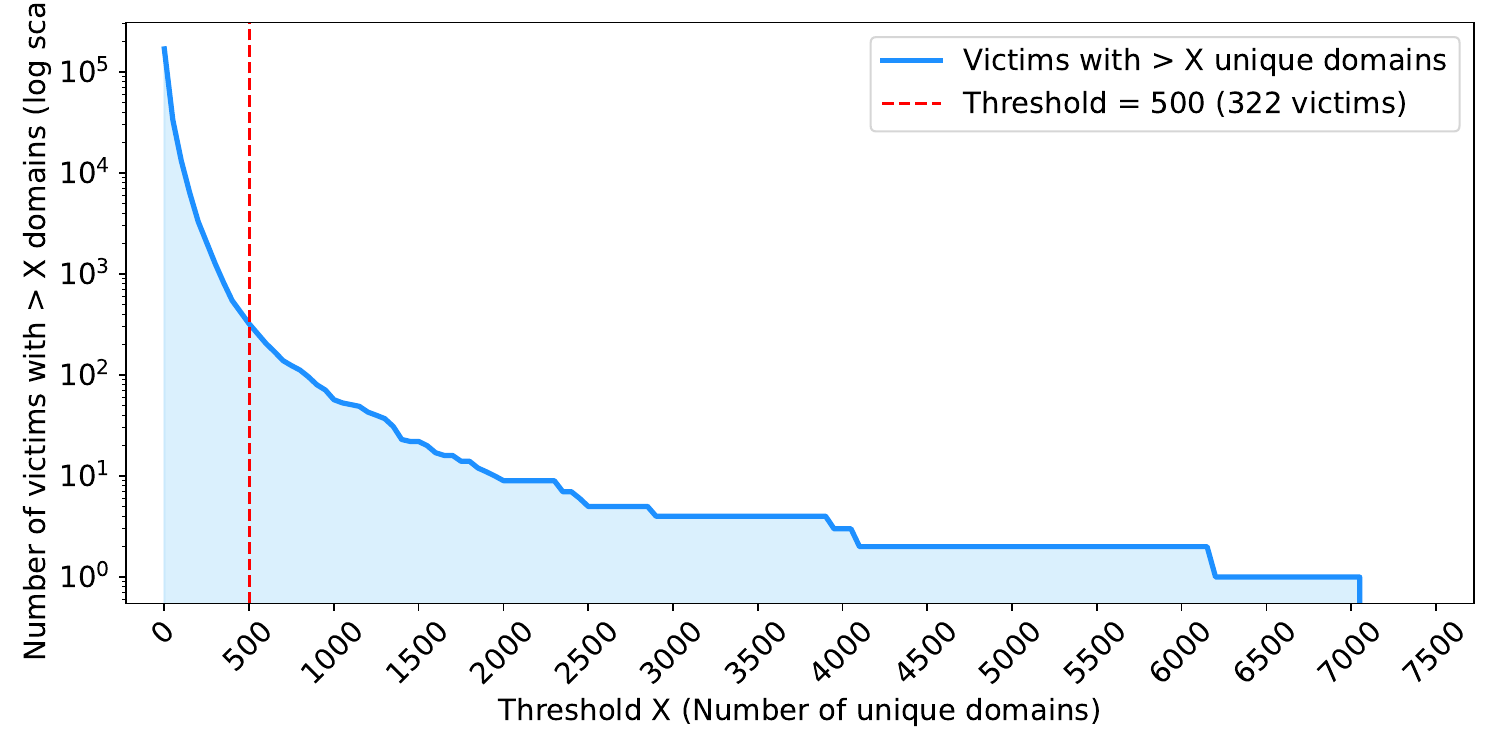}
        \caption{Number of victims with strictly more than $X$ unique
          domains, plotted on a logarithmic scale. The red dashed line
          marks the analytical threshold of $\ge 500$ unique domains
          ($n = 322$ victims).}\label{fig:histogram}
\end{figure}

\textbf{Method:}

Using the Cisco Talos API~\citep{talos_url_reputation}, we reduce each
URL where a victim stored credentials to its registered domain and
assign that domain to one of Talos's 106 domain
categories. We focus on two groups of categories: those identifying
recreational, personal, risky or illicit activities
(\autoref{tab:domain_categories}), and those indicating institutional
or professional activities (\qt{Government and Law}, \qt{Health and
  Medicine}, \qt{Education}, and \qt{Finance}).

\begin{figure}[!htb]
            \centering
        \includegraphics[width=1.0\linewidth]{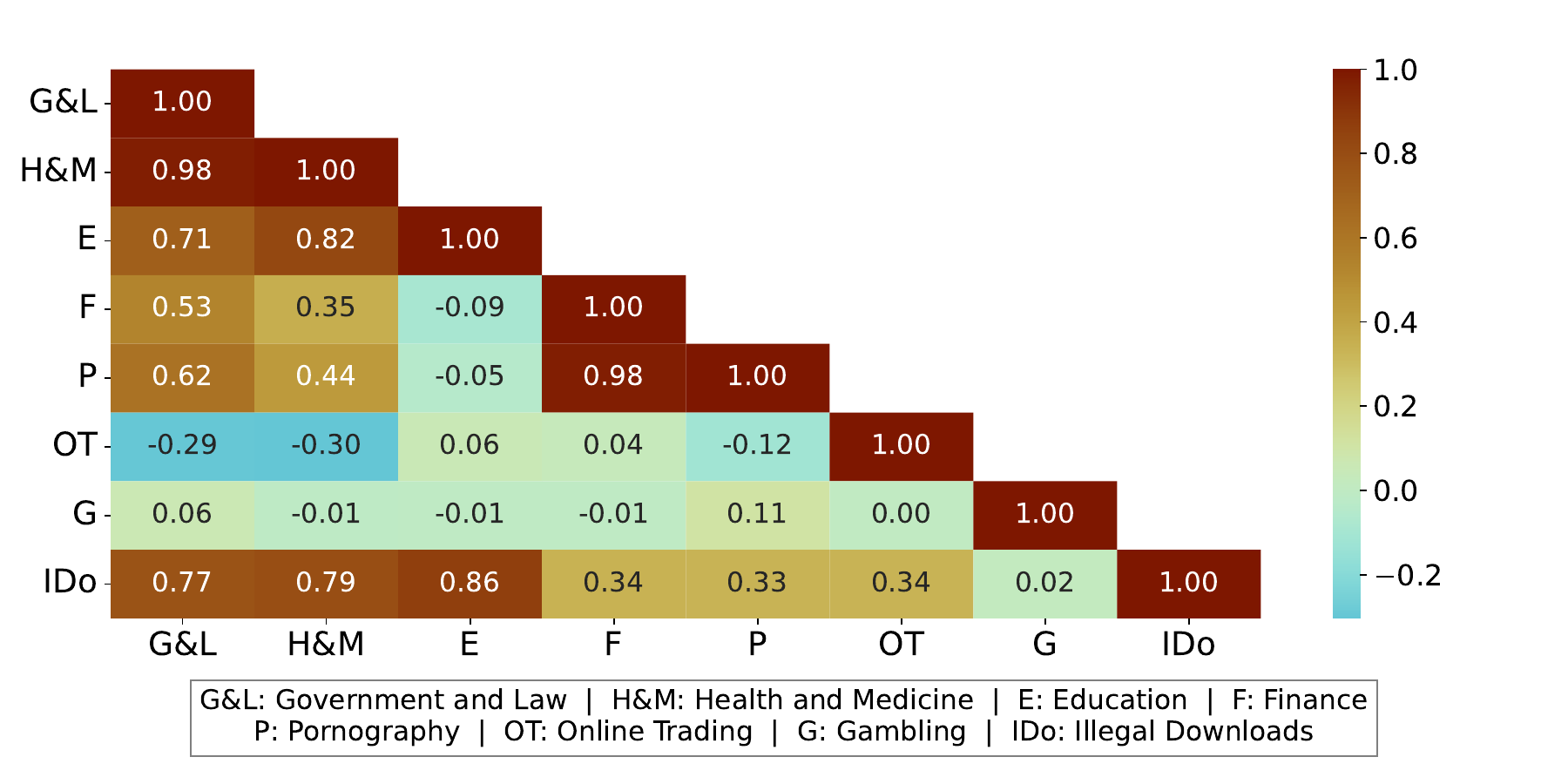}
\caption{Cosine similarity between the categories of credential-associated service usage, under
  strict filtering ($> 500$ domains per victim, $\ge 200$ victims per category; $n = 322$
  victims).}\label{fig:categories_correlation}
\end{figure}

Besides focusing on victims
with more than 500 domains, we retain only domain categories
with at least 200 victims, corresponding to 42 categories, and analyze
how these categories co-occur in victims' credential
logs. For each victim we normalize the number of domains in each category by that victim's total
number of domains. We reduce the resulting victim-by-category
matrices via SVD\@.
We retain the smallest number of components whose cumulative explained
variance reaches $85\,\%$, capturing the dominant structure while
discarding noise. The number of dimensions selected is $k=4$. Finally, we compute the
cosine similarity between the category columns
(see \autoref{fig:categories_correlation}).

\begin{table}[!htb]
  \centering \footnotesize
  \caption{List of categories associated with recreational, personal, risky or illicit services.}\label{tab:domain_categories}
  \rowcolors{1}{gray!10}{white}
  \begin{tabular}{@{}>{\raggedright\arraybackslash}p{0.30\columnwidth}
                  >{\raggedright\arraybackslash}p{0.30\columnwidth}
                  >{\raggedright\arraybackslash}p{0.30\columnwidth}@{}}
    \toprule
  Illegal downloads & Gambling & Pornography \\
  Weapons & Illegal drugs & Cannabis \\
  Cheating  \& plagiarism & Filter avoidance & Hate speech \\
  Cryptomining & Extreme & Illegal activities \\
  Adult & Online Trading & Alcohol \\
  Terrorism \& violent extremism & Child sexual abuse material & Hacking \\
    \bottomrule
  \end{tabular}
  \end{table}

\textbf{Results:} Among these victims, deviant activities intersect
directly with professional and work-related services on the same
machine (\autoref{fig:categories_correlation}).

Specifically, we observe a strong alignment among institutional and
professional usage: \qt{Government and Law} and \qt{Health and
  Medicine} exhibit an extremely high correlation ($0.98$), grouping
closely with \qt{Education} via robust correlations of $0.71$ and
$0.82$, respectively. Notably, \qt{Finance} and
\qt{Pornography} show an exceptionally strong correlation of
$0.98$ with each other, while both keep moderate positive
correlations with institutional sectors: \qt{Pornography} correlates
$0.62$ with \qt{Government and Law} and $0.44$ with \qt{Health and
  Medicine}, and \qt{Finance} $0.53$ and $0.35$ with the same two
categories. Furthermore, \qt{Illegal
  Downloads} maintains unexpectedly strong positive correlations with
these core institutional categories ($0.77$ with \qt{Government and
  Law}, $0.79$ with \qt{Health and Medicine}, and $0.86$ with
\qt{Education}), rather than standing apart as an isolated high-risk
cluster. Conversely, \qt{Online Trading} shows weak negative
correlations with institutional sectors ($-0.29$ with \qt{Government
  and Law} and $-0.30$ with \qt{Health and Medicine}), while
\qt{Gambling} is essentially uncorrelated with them ($0.06$ and
$-0.01$). The presence of this dual
usage demonstrates a critical intersection of risk, where infostealer
infections capture victims who combine personal or recreational risk
behaviors with critical professional activity, making the leaked
credentials exceptionally sensitive.

\begin{takeout}
    Analyzing victims with negligent credential retention practices
    reveals an overlapping
    exposure vector: personal and recreational activities intersect
    directly with professional services on the same machines, indicating
    that victims engaged in high-risk behaviors also store credentials for
    institutional services, rendering the leaked data highly sensitive.
\end{takeout}

\subsection{Critical domains and victims}\label{sec:special_domains}

RQ4 asks how often infostealers reach security-critical and other high-profile organizations.

\textbf{Method:}
We searched for victims who had authenticated to specific high-value domains, taking the eight Ivy League universities
(Brown, Columbia, Cornell, Dartmouth, Harvard, Pennsylvania, Princeton, and Yale) as an example of well-known institutions,
alongside security- and privacy-focused organizations and government bodies.
\autoref{tab:grouped_domain_counts} groups these domains by the kind of organization that runs
them, keeping commercial services separate from institutions whose members are affiliated with them.
These are illustrative cases rather than a representative sample of affected organizations;
the global spread of infections is reported in \autoref{sec:geography}.

\textbf{Results:}
This search surfaces compromised credentials across all eight Ivy League universities; technology and security-relevant organizations
including \breaktt{torproject.org} and \breaktt{linuxfoundation.org} (\autoref{tab:grouped_domain_counts});
and both \breaktt{interpol.int} and \breaktt{dhs.gov} (United States Department of Homeland Security) together with 32 other police domains.

\begin{table}[!htb]
  \centering
  \footnotesize
  \caption{Analyzing anecdotal evidence: compromised credentials for high-profile services, grouped by the kind of organization that runs the domain.}\label{tab:grouped_domain_counts}
  \rowcolors{2}{gray!10}{white}
  \begin{tabular}{p{2.5cm}rp{2.5cm}r}
  \toprule
  \textbf{Domain} & \textbf{Count} & \textbf{Domain} & \textbf{Count} \\
  \midrule
  \multicolumn{4}{@{}l}{\emph{Universities and research institutes}} \\
mit.edu & 4,158 & yale.edu & 35 \\
harvard.edu & 129 & ox.ac.uk & 29 \\
stanford.edu & 118 & columbia.edu & 28 \\
berkeley.edu & 88 & mpg.de & 27 \\
cornell.edu & 70 & uchicago.edu & 17 \\
upenn.edu & 63 & princeton.edu & 10 \\
cam.ac.uk & 48 & brown.edu & 6 \\
cmu.edu & 35 & dartmouth.edu & 3 \\
\midrule
\multicolumn{4}{@{}l}{\emph{Scholarly publishers and preprint servers}} \\
springernature.com & 82 & arxiv.org & 22 \\
nature.com & 50 & biorxiv.org & 2 \\
\midrule
\multicolumn{4}{@{}l}{\emph{Professional and standards bodies}} \\
ieee.org & 448 & usenix.org & 1 \\
acm.org & 53 &  &  \\
\midrule
\multicolumn{4}{@{}l}{\emph{Open-source and privacy foundations}} \\
linuxfoundation.org & 19 & torproject.org & 2 \\
gnu.org & 2 &  &  \\
\midrule
\multicolumn{4}{@{}l}{\emph{Commercial services}} \\
openai.com & 16,641 & sans.org & 49 \\
\midrule
\multicolumn{4}{@{}l}{\emph{Government and law enforcement}} \\
dhs.gov & 42 & interpol.int & 2 \\

  \bottomrule
 \end{tabular}
\end{table}

Many of the compromised services suggest
that misuse of the harvested credentials could enable access to sensitive organizational resources.
In particular, compromised organizational email accounts may serve as an entry point to broader systems by enabling password reset mechanisms across connected services.
\autoref{tab:illustrative_examples} provides illustrative examples of such compromised domains.
Following responsible-disclosure practice, we report these at the registered-domain level rather than
naming the exact endpoints that were reached; the released dataset applies the same reduction.
Unauthorized access to such infrastructure may have serious implications for institutional integrity, confidentiality, and the continuity of operations.

\begin{table}[!htb]
\centering
\scriptsize
\caption{Examples of compromised sensitive organizational domains.}\label{tab:illustrative_examples}
\rowcolors{2}{gray!10}{white}
\begin{tabular}{@{}>{\raggedright\arraybackslash}p{0.43\columnwidth}
                >{\raggedright\arraybackslash}p{0.49\columnwidth}@{}}
\toprule
\textbf{Domain} & \textbf{Description} \\
\midrule
\breaktt{army.mil.ph} & Philippine Army, email system \\
\breaktt{prefecturanaval.gov.ar} & Argentine Naval server \\
\breaktt{tni-au.mil.id} & Indonesian Air Force Academy \\ %
\breaktt{osd.mil} & U.S. Department of Defense \\
\breaktt{army.mil} & U.S. Army login and authentication portals \\
\breaktt{crmforce.mil} & The U.S. Air Force Academy \\
\breaktt{tavanir.org.ir} & Iranian electric grid company \\
\breaktt{who.int} & WHO intranet and webmail systems \\
\breaktt{navy.mil.ph} & Philippine Navy email server \\
\breaktt{boeing.com} & Boeing secure login platform \\
\breaktt{af.mil} & U.S. Air Force personnel and engineering portals \\
\breaktt{jaf.mil.jo} & Jordanian Armed Forces email server \\
\bottomrule
\end{tabular}
\end{table}

\begin{takeout}
  Stolen credentials reach highly sensitive organizations, including all eight Ivy League universities, government
  and military portals, Interpol, and more than 30 police forces, any of which could give attackers a foothold
  inside the institution.
\end{takeout}

\subsection{IP addresses and administration access}\label{sec:ip_addresses} %

The data contains login information for well-known platforms accessed via standard domain names, but victims also
authenticated directly via IP addresses (e.g., \breaktt{https://x.x.x.x/admin/}), which often points to
private systems such as admin panels or development environments, as well as to domains on non-public top-level
domains used only within private networks or VPN infrastructures. The logs also capture more than passwords:
private VPN configurations, digital certificates, SSH and PGP private keys, and other authentication tokens.
An adversary holding such credentials and URLs could pivot across systems along three vectors: bypassing security
controls to impersonate users and enable lateral movement, privilege escalation, and persistence; authenticating
directly to an IP address to reach non-public systems; and authenticating to remote-access
services (e.g., TeamViewer) or developer platforms to seize full control of a host or entry into development
pipelines and cloud environments. The recent compromise of Ingram Micro~\citep{cybernews_ingram_micro} shows how
costly exposed VPN access to critical infrastructure can be.

\textbf{Method:}
We compile every entry where a victim authenticated directly to an IP address and geolocate those we can. For these
hosts we query Shodan for open ports, running services, operating systems, and known vulnerabilities, which an
adversary could use to compromise them further (see \autoref{tbl:shodan_stats2}). We also identify credentials for
remote-access and developer-oriented services, and we extract the domains that use internal-only top-level domains (TLDs).

\begin{figure}[!htb]
  \centering
  \includegraphics[width=\columnwidth]{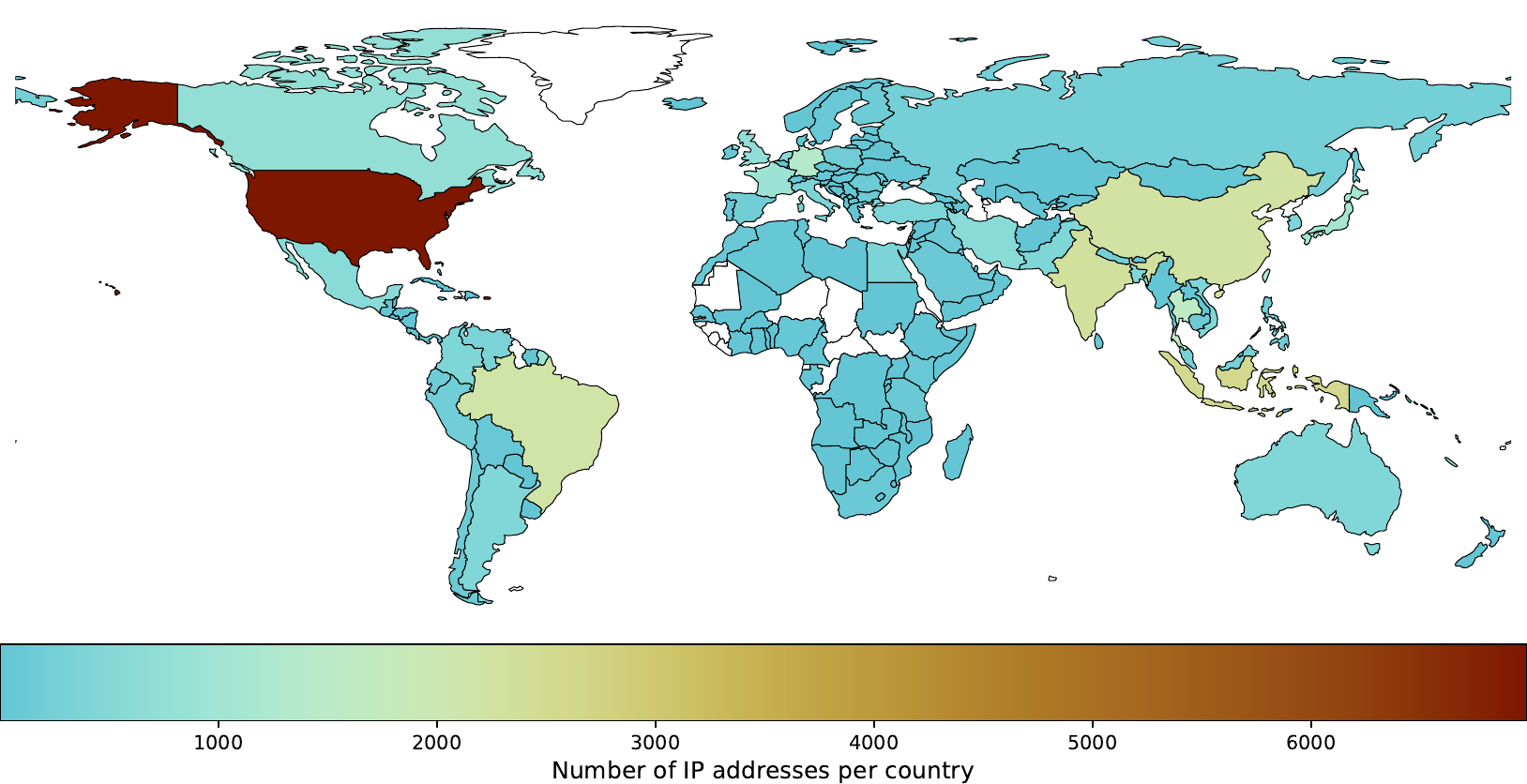}
  \caption{Direct IP addresses victims have authenticated per country.}\label{fig:ext_ips_map}
\end{figure}

\textbf{Results:}
Turning these vectors into measurements, we find that victims authenticated to 50,219 direct IP addresses; of these, 38,760 could be geolocated and are mapped in \autoref{fig:ext_ips_map}.
Some of these services require multi-factor authentication, e.g., \breaktt{github.com};
however, since the victim's system is compromised, an adversary could extract the necessary cookies and
authenticate, bypassing the multi-factor authentication.
Also, we find leaked access to popular software development services, including
\breaktt{github.com},
\breaktt{aws.amazon.com},
\breaktt{teamviewer.com},
and \breaktt{docker.com}.
Beyond that, we identify domains that do not use standard or publicly accepted TLDs, but instead rely on internal TLDs such as \breaktt{.local}, \breaktt{.lan}, and similar.
In total, we extract 4,385 such domains, counting each domain once regardless of how many victims reached it
and excluding \breaktt{.onion} addresses, which are Tor services rather than internal hosts.
Examples include addresses like \breaktt{http://army.lan}.
Many of these domains appear to reference internal administrative interfaces, databases, or corporate services intended for employee access within closed organizational environments.

\begin{figure}[!htb]
  \includegraphics[width=.8\columnwidth]{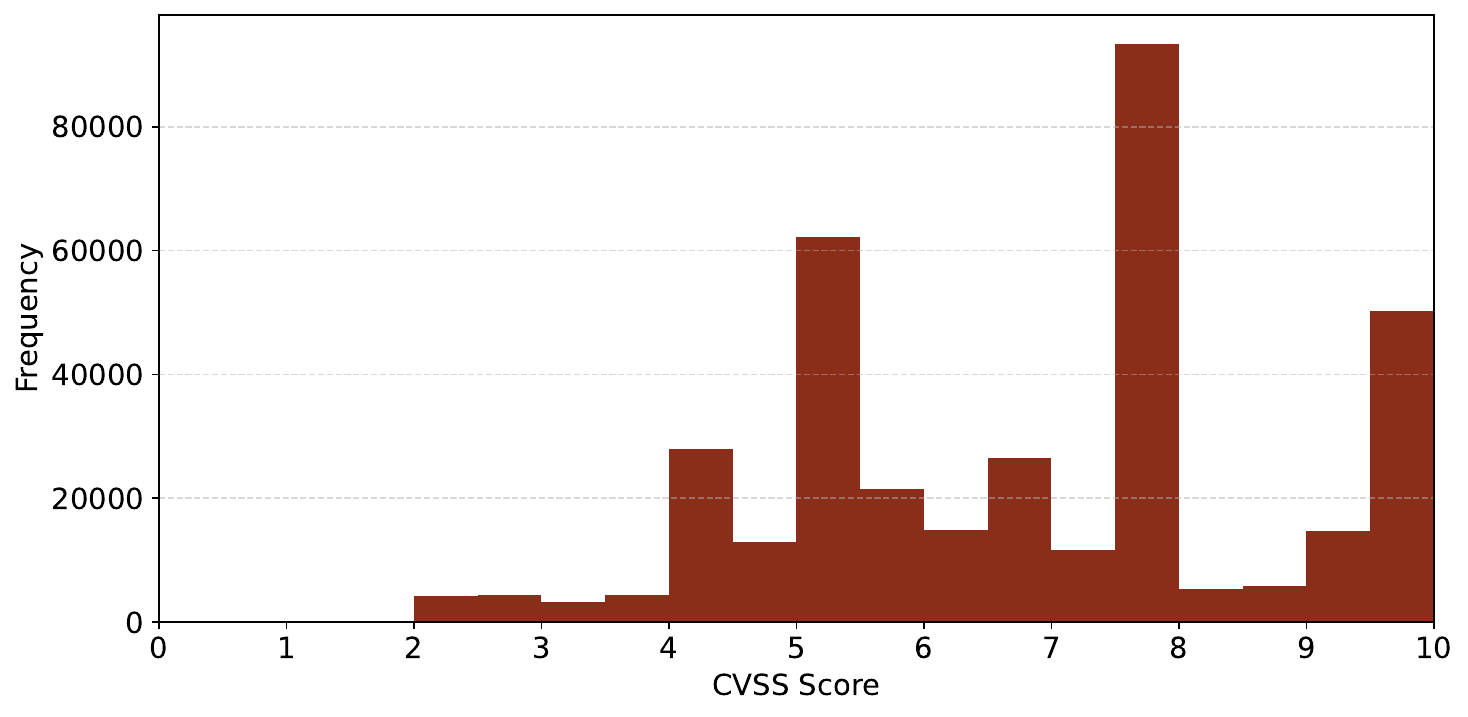}
  \caption{CVSS distribution of vulnerabilities.}\label{fig:cvss}
\end{figure}

From the 38,760 external IPs, we found 13,162 scanned by Shodan.
The operating system and open ports statistics are provided in \autoref{tbl:os}
and \autoref{tbl:ports}, respectively.  Because
victims accessed these systems through a browser, most of the systems expose a common
port used by a web server, e.g., 80, 443,
and 8080; %
however, other ports commonly used by SSH (port 22, n = 4,256) and File Transfer Protocol (FTP, port 21, n = 1,199), %
along with MySQL (port 3306, n = 1,048) and Remote Desktop Protocol (RDP, port 3389, n = 852), are also exposed. %
Additionally, Shodan reports known vulnerabilities for many of these systems, whose Common Vulnerability Scoring System (CVSS) score distribution is
illustrated in \autoref{fig:cvss}, implying that many of these systems are potentially exposed to severe threats.

\begin{table}[!htb]
  \centering
  \caption{Shodan statistics.}\label{tbl:shodan_stats2}
  \scriptsize
  \begin{subtable}[t]{0.45\textwidth}
    \centering
    \caption{Most common ports.}\label{tbl:ports}
    \rowcolors{2}{gray!10}{white}
    \begin{tabular}{lrlr}
      \toprule
      \textbf{Port} & \textbf{Count} & \textbf{Port} & \textbf{Count} \\
      \midrule
      80   & 7,088 & 53   & 1,165 \\ %
      443  & 5,526 & 3306 & 1,048 \\ %
      22   & 4,256 & 2000 & 959   \\ %
      8080 & 1,231 & 3389 & 852   \\ %
      21   & 1,199 & 161  & 750   \\ %
      \bottomrule
    \end{tabular}
  \end{subtable}\hfill
  \begin{subtable}[t]{0.45\textwidth}
    \centering
    \caption{Most common operating systems.}\label{tbl:os}
    \rowcolors{2}{gray!10}{white}
    \begin{tabular}{p{0.65\linewidth}r}
      \toprule
      \textbf{Operating system} & \textbf{Count} \\
      \midrule
      Linux & 1,197 \\
      Windows & 995 \\
      Ubuntu & 597 \\
      Windows Server 2022 (build 10.0.20348) & 89 \\ %
      Windows (build 10.0.17763) & 64 \\ %
      Debian & 64 \\
      Windows (build 10.0.14393) & 56 \\ %
      Windows (build 6.3.9600) & 42 \\ %
      Windows Server 2012 R2 (build 6.3.9600) & 40 \\ %
      Windows (build 10.0.19041) & 37 \\ %
      \bottomrule
    \end{tabular}
  \end{subtable}
\end{table}

Moreover, there are 1,233 URLs with leaked credentials pointing to intranet services, 122 of which belong to
\texttt{gov} domains. Likewise, there are 484 leaked credentials to governmental mail services.

\begin{takeout}
  Beyond website logins, victims stored credentials for 50,219 IP-addressed hosts and 4,385 internal-only domains,
  including admin panels and exposed SSH, RDP, and database services that an adversary could use to pivot into internal networks.
\end{takeout}

\subsection{Hacking and carding}\label{sec:crime} %

So far we have treated victims purely as targets. To address RQ5, we examine the illegal activity that can be inferred
from the logs, asking whether some victims appear to be involved in cybercrime themselves.

\textbf{Method:}
Carding forums are online communities dedicated to trading stolen payment-card data and associated fraud services,
whereas hacking forums trade exploits, compromised accounts, and other offensive cyber services.
While crawling the 404 forums introduced in \autoref{sec:datacollection}, we found that 205 of them, such as \texttt{cardforum.cc}, hosted carding and hacking material behind a login.
We used this sublist of 205 forums to count how many infostealer victims authenticated into carding and hacking communities.

The Tor Browser enables users to access anonymous onion websites.
Infostealer malware captures this usage data, providing us with a mechanism to study those who have authenticated into anonymous onion websites on Tor.
We analyze victims who have authenticated to at least one onion website using a Tor Browser.

\begin{figure}[!htb]
    \includegraphics[width=1.0\linewidth]{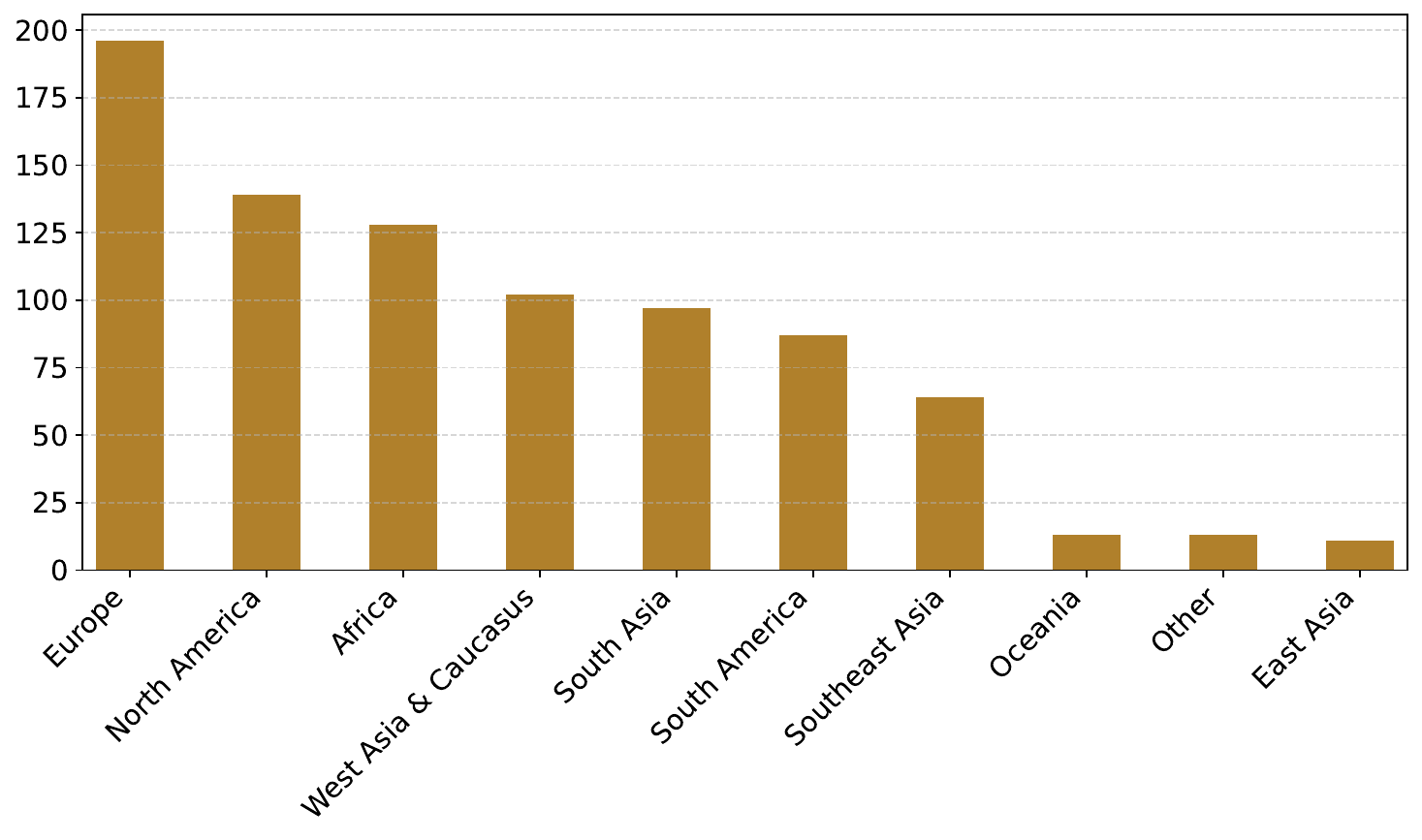}
    \caption{
    Region-specific breakdown of hacking and carding forum victims.}\label{fig:region_bar}
\end{figure}

\textbf{Results:}
We identify 850 victims who authenticated to at least one carding or hacking forum site.
A further 188 victims authenticated to onion sites on Tor, commonly to illegal marketplaces.
The geographic distribution of the carding and hacking forum victims is shown in \autoref{fig:region_bar}.
While it does not demonstrate a large population, it does show the ability to identify risky and possibly illicit activities within the dataset.

\begin{takeout}
  A small subset of victims appear to be involved in cybercrime themselves: 850 authenticated to
  carding and hacking forums and 188 to Tor onion sites, commonly illegal marketplaces.
\end{takeout}

\subsection{Financial service exposure}\label{sec:financial}

RQ6 asks how exposed victims' high-value commercial services are. We begin with financial
platforms, where a compromised credential translates most directly into monetary loss. Financial harm can also
arrive indirectly, for example when a payment card linked to a Google, Amazon, or Apple account enables an adversary to
make unauthorized purchases, but this indirect impact is hard to quantify.

\textbf{Method:}
A victim is financially exposed if the logs hold a credential for a domain where an account keeps or moves money.
We used the public API of the Federal Deposit Insurance Corporation (FDIC)~\citep{fdic_bank_data} and the bank
membership list of the Euro Banking Association~\citep{eba_members} to assemble a list of 4,221 bank domains. We took
payment processors, exchanges, brokerages, and mining pools from the Talos categorization introduced in
\autoref{sec:victim-authentication}, using its Finance, Cryptocurrency, Online Trading, and Cryptomining categories.
The two sources give 31,414 financial domains in total, every one of them holding a balance or a payment instrument
that a stolen credential can reach.
We report how many victims hold at least one such credential and how many distinct financial domains each stored.

\textbf{Results:}
Infostealers collected financial credentials from 73,782 victims, 43.33\,\% of the dataset, who stored on average
4.84 distinct financial domains each. \autoref{tbl:fin} lists the 20 domains held by the most victims.
The exposure sits with payment and cryptocurrency services rather than with banks. PayPal alone accounts for
39.1\,\% of affected victims, while only 4,152, or 2.44\,\% of the dataset, hold a credential for any of the 4,221
bank domains. The two bank sources cover only US and European institutions, so banks elsewhere reach the analysis
through the Talos categorization instead, \texttt{caixa.gov.br}, \texttt{hdfcbank.com}, and \texttt{nubank.com.br}
among them. Because the bank list is regional, the 2.44\,\% is a lower bound on bank exposure
rather than a global rate, though it remains an order of magnitude below the payment platforms.
This suggests that despite the wide adoption of e-banking, victims infrequently store their credentials
for banking domains. In the case of European financial institutions, this could reflect the stringent authentication requirements established by the
European Union's second Payment Services Directive (PSD2)~\citep{PSD2}.
As underlined by infostealer attacks, impersonation is a critical matter for e-payments,
and for this reason PSD2 is currently being updated in part to address spoofing attacks in the context of a proposed PSD3.

\begin{table}[!htb]
\centering
\caption{The 20 financial domains held by the most victims. Each victim is counted once per domain, so a
victim with accounts at several of them appears in several rows.}\label{tbl:fin}
\scriptsize
\rowcolors{2}{gray!10}{white}
\begin{tabular}{lrlr}
\toprule
\textbf{Domain} & \textbf{Victims} & \textbf{Domain} & \textbf{Victims}\\
\midrule
paypal.com & 28,876 & iqoption.com & 1,550 \\
binance.com & 5,867 & onlinesbi.sbi & 1,386 \\
coinbase.com & 4,062 & olymptrade.com & 1,346 \\
caixa.gov.br & 3,907 & hdfcbank.com & 1,338 \\
payoneer.com & 3,146 & sandbox.game & 1,319 \\
nicehash.com & 2,235 & qxbroker.com & 1,280 \\
paysafecard.com & 2,095 & blockchain.com & 1,256 \\
exness.com & 2,004 & faucetpay.io & 1,242 \\
stripe.com & 1,868 & intuit.com & 1,149 \\
wise.com & 1,728 & nubank.com.br & 1,090 \\
\bottomrule
\end{tabular}
\end{table}

\begin{takeout}
Infostealers exposed financial credentials for 73,782 victims, 43.33\,\% of the dataset, concentrated in payment
platforms (led by PayPal) and cryptocurrency services rather than traditional banks, which reach only 2.44\,\%. The pattern tracks authentication strength: banks enforce
regulated strong authentication such as PSD2, so the exposure shifts to payment and crypto services that often do
not, where browser-stored credentials leave accounts directly at risk.
\end{takeout}

\subsection{Communication and privacy-oriented services}\label{sec:comm_privac}

The other commercial category under RQ6 is communication and privacy: victims also expose the collaboration platforms
they use for work and the services they rely on to protect their privacy.

\textbf{Method:}
We compare communication and privacy-oriented services in our dataset, including both mainstream collaboration platforms
and privacy-focused alternatives. For mainstream platforms, we search for Zoom, Webex, and Slack as widely used enterprise collaboration tools.
For privacy-focused services, we search \breaktt{tuta.com} and \breaktt{tutanota.com} (Tuta Mail),
\breaktt{proton.me} and \breaktt{protonmail.com} (Proton Mail), and VPN services.
This comparison allows us to assess the relative prevalence of privacy-oriented services among infostealer victims.

\textbf{Results:}
Zoom (n = 11,422), Webex (n = 1,275), and Slack (n = 932) are common collaboration services in the dataset.
Proton Mail appears in 1,963 victim profiles (\rounding{1963/170298*100}\,\% of our sample).
For comparison, Proton reports approximately 100 million accounts worldwide\footnote{\url{https://proton.me/blog/proton-100-million-accounts}},
representing roughly \rounding{100/5560*100}\,\% of the global internet user base~\citep{datareportal2025}.
The similar privacy-driven email service Tuta is considerably less popular (n = 177).
Furthermore, if we seek the pattern \qt{VPN} in the list of accessed URLs, we can identify victims using a VPN
service (n = 9,929), with particular relevance for those using services such as ProtonVPN, NordVPN, or SurfSharkVPN,
as these VPN services are categorized as privacy-respectful~\citep{privacytools}.

\begin{takeout}
  Privacy tooling is not enough on its own to avoid infostealers. Users of privacy-focused email and VPNs appear
  among victims, suggesting these services offer limited
  protection against infection, which is unsurprising given that infostealers compromise the endpoint rather than
  the network.
\end{takeout}

These results collectively reveal an intersection of risk: infostealer infections indiscriminately capture a large
global population (\autoref{sec:geography}, \autoref{sec:os}), yet the resulting data facilitates future targeted
attacks against critical infrastructure (\autoref{sec:ip_addresses}), law enforcement and other sensitive
institutions (\autoref{sec:special_domains}), and financial services (\autoref{sec:financial}).
Gaming and entertainment services are strongly overrepresented among victims
(\autoref{sec:representativeness}), who demonstrate weak baseline security practices across both personal and
professional domains (\autoref{sec:victim-authentication}, \autoref{sec:password_analysis}).

\subsection{Differences between macOS and Windows victims}\label{sec:family_strata}

\textbf{Method:}
Our logs carry no infostealer family label for individual victims, because the attribution
in~\autoref{tbl:collected_logs} applies to a collection of logs rather than to a single record.
The operating system still separates the families, since a macOS host cannot have been infected by
the Windows-only families and Atomic macOS Stealer reaches no Windows host, so comparing the two
platforms is the closest our data gets to comparing families.
Separately for the 4,523 macOS and the 151,161 Windows victims, we recompute the median number of
domains a victim stored credentials for, the share of victims holding credentials in the deviant and
institutional category groups of~\autoref{sec:victim-authentication}, the share reaching a direct IP
address or an internal TLD (\autoref{sec:ip_addresses}), and the password reuse rate
of~\autoref{sec:password_analysis}.
We add one group, workplace categories that credentials reach more often through work than through
personal use: Organizational Email, SaaS and B2B, Online Meetings, Business and Industry, and
Professional Networking.
Every measurement except the medians is the percentage of victims who stored credentials for at
least one domain of that kind, or who reused at least one password.
We count all victims of each platform, not only the 322 who authenticated to more than 500 domains
in~\autoref{sec:victim-authentication}.
Percentages carry Wilson score intervals and the medians percentile bootstrap intervals.
The macOS victims live in different countries and store more credentials than the Windows victims,
and either difference could produce a gap on its own.
We therefore group the victims by country, by how many domains they stored, and by both together,
take the macOS percentage inside each group, and average those percentages using the Windows group
sizes as weights.
The result is the percentage the macOS victims would show if they were spread over the groups like
the Windows victims (direct standardization).

\textbf{Results:}
Every measurement holds on both platforms in the same order, but the magnitudes differ
(\autoref{tbl:os_strata}).
The macOS victims stored credentials for more domains, were more exposed on institutional, financial
and workplace services, and were about twice as likely to store credentials for an internal host,
yet they reused passwords less often.
None of the three adjustments moves a macOS percentage by more than 4.9 points, so neither
geography nor the volume of stored credentials accounts for these differences.
The deviant category moves the most, and its gap to Windows narrows to 2.7 points.

\begin{table}[!htb]
    \centering
    \caption{Victim exposure and password reuse by operating system, with 95\,\% intervals.
    The category rows give the percentage of victims who stored credentials for at least one such
    domain, the last row the percentage who reused at least one password.
    The three middle columns repeat the macOS percentage as if the macOS victims matched the
    Windows victims on country, on stored domains per victim, and on both at once.
    The joint adjustment covers 92.4\,\% of the Windows victims, the other two 98.2\,\% and
    100\,\%. The uncovered Windows victims fall in country and volume combinations that hold no
    macOS victim, leaving no macOS rate to reweight, so the adjusted columns are renormalized over
    the covered weight.}\label{tbl:os_strata}
    \scriptsize
    \setlength{\tabcolsep}{3.5pt}
    \rowcolors{3}{gray!10}{white}
\begin{tabular}{lrrrrr}
\toprule
 & \multicolumn{4}{c}{\textbf{macOS}} & \\
\cmidrule(lr){2-5}
\textbf{Measurement} & \textbf{Observed} & \textbf{Country} & \textbf{Volume} &
\textbf{Both} & \textbf{Windows}\\
\midrule
Domains (median) & 45 [42--49] & & & & 36 [35--36] \\
Deviant & 49.0 [47.5--50.4] & 48.8 & 46.4 & 44.1 & 41.4 [41.2--41.7] \\
Institutional & 78.4 [77.1--79.5] & 79.0 & 77.5 & 77.6 & 67.7 [67.5--68.0] \\
Finance & 54.9 [53.4--56.3] & 54.7 & 52.3 & 50.7 & 38.0 [37.8--38.2] \\
Workplace & 79.5 [78.3--80.7] & 79.0 & 78.9 & 76.8 & 65.7 [65.5--66.0] \\
Direct IP address & 37.8 [36.4--39.2] & 39.8 & 35.3 & 35.3 & 27.4 [27.2--27.6] \\
Internal TLD & 5.7 [5.0--6.4] & 5.7 & 5.1 & 4.6 & 2.9 [2.8--2.9] \\
Reused a password & 84.9 [83.9--86.0] & 85.6 & 85.0 & 85.0 & 89.1 [88.9--89.3] \\
\bottomrule
\end{tabular}
\end{table}

\begin{takeout}
  Every finding holds on both platforms, but not at the same magnitude. A rate measured on one
  platform should be read as specific to that platform.
\end{takeout}

The next section asks whether these victims are also exposed to other forms of cybercrime.
\section{Revictimization}\label{sec:revictimization}

An infostealer infection is rarely a victim's only security problem. To address RQ3, we examine whether victims have also
fallen for phishing, what their passwords reveal about their cyber hygiene, and whether organizations in our dataset later
appear as ransomware victims.

\subsection{Revictimization through phishing}

\textbf{Method:}
We identify victims who have authenticated to phishing sites.
Suspicious TLDs such as \texttt{.shop}, \texttt{.top}, and \texttt{.xyz} accounted for roughly 37\,\% of cybercrime domains
reported between September 2023 and August 2024~\citep{Interisle}, so we query VirusTotal about the domains with these TLDs
in our dataset and add the phishing domains the Anti-Phishing Working Group (APWG) published for December 2024 and January 2025.
A domain counts as phishing if at least one VirusTotal engine flagged it as malicious or if it appears in the APWG
feed. Both labels describe a domain as seen at or after our collection, and the logs do not record when a credential
was saved, so we count victims who stored credentials for domains labelled as phishing rather than victims phished
at that moment. Covering three TLDs and two months of APWG
reports, the result is a lower bound.

\textbf{Results:}
The set comprises 20,597 domains (11,881 \texttt{.xyz}, 5,767 \texttt{.top}, and 2,949 \texttt{.shop}).
VirusTotal had scanned almost all of them (only 73 were not), and 465 carried at least one malicious detection.
We find that 1,037 victims disclosed credentials to such fraudulent websites, most of them reported for phishing,
and 143 appear in the APWG set. Four victims appear in both, giving 1,176 distinct phishing victims.

\begin{takeout}
  Infostealer victims are revictimized through other fraud channels: at least 1,176 had separately
  disclosed credentials to known phishing sites.
\end{takeout}

\subsection{Digital hygiene}\label{sec:password_analysis}

Good digital hygiene means applying least privilege, separating work from personal activity, using strong and unique
passwords for every service, and keeping the operating system and applications patched.

\textbf{Method:}
We measure credential hygiene directly from the passwords and usernames stored in each victim's logs. For every
victim we count the stored passwords and usernames, how many are distinct and how many are unique, and whether a
username is reused as a password. Distinct counts the different values a victim stored, so a password saved three
times contributes one, while unique counts only the values stored exactly once, so that same password contributes
none. We use unique in two senses and label every figure below with the one it follows:
within-victim unique means a value that occurs exactly once in that victim's own logs, and dataset-wide unique means
a value that occurs exactly once across all victims pooled together. We then aggregate these counts across all
victims and compare the resulting within-victim rates against prior measurements of user-chosen credentials, which
are also reported per user.

\textbf{Results:}
Victims rarely meet these standards, starting with their passwords. Our dataset contains a total of 28,026,689
passwords, of which only 1,638,018
(\rounding{1638018/28026689*100}\,\%) are dataset-wide unique.
We find that \rounding{151547/170298*100}\,\% (n = 151,547 of 170,298)
of victims reuse passwords and \rounding{148036/170298*100}\,\% (n
= 148,036 of 170,298) reuse usernames. Additionally,
\rounding{70050/170298*100}\,\% (n = 70,050 of 170,298) of victims
use their usernames as passwords.

The average victim has 164.57 password entries.
Of these, only \rounding{29.72/164.57*100}\,\% (n = 29.72 of 164.57) are distinct,
while the remaining \rounding{134.85/164.57*100}\,\% (n = 134.85 of 164.57) are repetitions.
Furthermore, only \rounding{12.20/164.57*100}\,\% (n = 12.20 of 164.57) are within-victim unique,
indicating that even among distinct passwords, many are reused.
The most common password is reused 43.30 times on average.

Similarly, the average victim has 162.20 usernames,
of which \rounding{31.52/162.20*100}\,\% (n = 31.52 of 162.20) are distinct
and \rounding{13.09/162.20*100}\,\% (n = 13.09 of 162.20) are within-victim unique,
while the remaining \rounding{130.67/162.20*100}\,\% (n = 130.67 of 162.20) are duplicates.
The most common username is reused 38.95 times on average.

These observations are consistent with prior work~\citep{DBLP:journals/ijinfoman/BangLBA12},
which also finds that users rely on a small set of credentials reused across many accounts,
reporting similarly low proportions of within-victim unique passwords and usernames.
In particular, they report that only about \rounding{4.7/105.7*100}\,\% of passwords and \rounding{6.6/105.7*100}\,\% of usernames are unique per user,
compared to \rounding{12.20/164.57*100}\,\% and \rounding{13.09/162.20*100}\,\% in our dataset. Both sides of this
comparison use the within-victim definition; the dataset-wide figure above is not comparable to it.
Moreover, victims who stored credentials on phishing-flagged domains are a negligible share of the dataset
(1,176 of 170,298), so credentials saved on such sites cannot materially affect these reuse rates.

\begin{figure}[!htb]
  \includegraphics[width=\columnwidth]{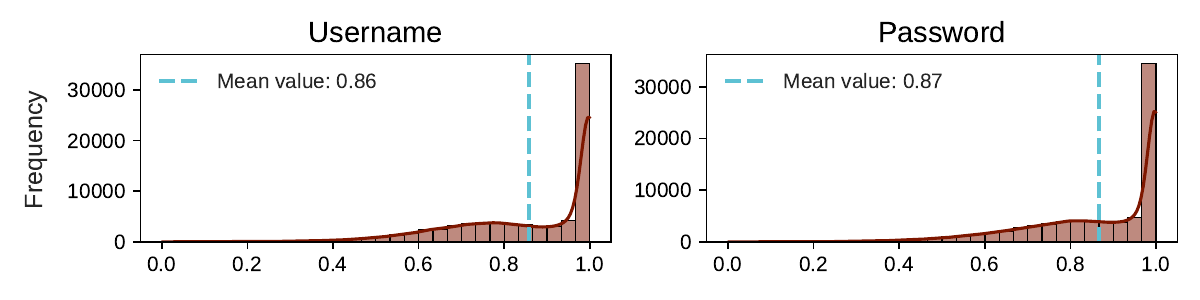}
  \caption{Statistics on credential leakage, filtered to include only victims with 50 or more credentials.}\label{fig:credentials_metrics}
\end{figure}

\autoref{fig:credentials_metrics} shows the repetition rate of usernames and passwords for victims with 50 or more credentials.
On average, 86\,\% of usernames and 87\,\% of passwords are reused.
Across the whole dataset, the single most common password accounts for 667,386 entries (\rounding{667386/28026689*100}\,\% of all passwords).

As shown in \autoref{fig:categories_correlation}, victims access work or critical services alongside recreational
and even illegal content on the same device, consistent with surveys on remote workers using work devices for
personal activities~\citep{ESET_2024,DBLP:conf/hci/NurseWCPBK21}, underscoring the need for effective
cyberawareness training~\citep{nwankpa2023}.

\begin{takeout}
  Victims show severe digital hygiene failures: \rounding{151547/170298*100}\,\% reuse passwords, only
  \rounding{1638018/28026689*100}\,\% of all passwords are unique, and \rounding{70050/170298*100}\,\% use a username
  as a password. A single victim's log typically holds both personal accounts and work or critical-service logins,
  so one compromise exposes both at once.
\end{takeout}

\subsection{Links to ransomware victims}

Beyond exploiting software vulnerabilities, ransomware groups such as LockBit often lure employees into leaking
credentials to gain an initial foothold in a target organization. Harvesting credentials from infostealer logs is
therefore a plausible shortcut to that foothold, and we measure how far the two victim populations overlap.

\textbf{Method:}
We collected all domains with leaked credentials in our dataset and all victims listed in the
\texttt{Ransomware.live} database~\citep{ransomwarelive}, keeping only the ransomware victims whose entry included a
URL we could match against, so the monthly totals in \autoref{tbl:ransomware} count only the victims we could resolve
to a domain. A match requires the domain in the Ransomware.live entry to equal a domain for which our dataset holds
credentials, and we inspected every candidate by hand to reject partial string matches. Because initial access, ransomware deployment, and the eventual publication of stolen
data can be separated by long and variable gaps, we widen the window and study ransomware victims from October
2024 to the end of May 2025. We report the triaged results in aggregate, counting only whether leaked credentials
exist for an organization, not how many.

\textbf{Results:}
The existence of leaked credentials for an organization does not imply that the organization will be compromised, as the
credentials might not provide elevated access to the system but rather harm the individual victim. For instance, credentials
belonging to a bank's or an e-shop's customers most likely harm those customers rather than the organization itself.
Nevertheless, the leaked credentials in our dataset are not associated solely with domains that are
expected to have many unprivileged users. Some of the matched domains are webmail and remote-access services, where a
compromised account can serve an adversary directly, for instance by abusing an employee's mailbox to induce colleagues
to open malicious files.
As shown in \autoref{tbl:ransomware}, leaked credentials appear across all months in the observation window,
including months before our January~2025 collection. We note that 54 matches occur in February~2025,
the peak month, which also coincides with the highest number of reported ransomware victims (n = 654).
A log records when we collected it, not when the credentials were stolen or when they began circulating, so we
cannot place a given exposure before or after the ransomware event it matches. The overlap is therefore evidence of
a shared victim population rather than of initial access.

\begin{table}[!htb]
  \centering
  \caption{Identified ransomware victims.}\label{tbl:ransomware}
  \scriptsize
  \rowcolors{2}{gray!10}{white}
  \begin{tabular}{lrr}
    \toprule
    \textbf{Month} &\textbf{\# Ransomware victims' URLs} &\textbf{Identified Leaked credentials}\\
    \midrule
    Oct 2024 & 238 & 23\\
    Nov 2024 & 286 & 21\\
    Dec 2024 & 311 & 43\\
    Jan 2025 & 369 & 48\\
    Feb 2025 & 654 & 54\\
    Mar 2025 & 356 & 31\\
    Apr 2025 & 139 & 5\\
    May 2025 & 228 & 11\\
    \bottomrule
  \end{tabular}
\end{table}

\begin{takeout}
  Organizations listed as ransomware victims frequently also appear in infostealer logs as domains with exposed
  employee credentials. We measure the overlap between the two populations, not initial access, but an overlap this
  size is reason enough to treat infostealer exposure as a warning sign: reset the affected credentials and enforce
  multi-factor authentication on email and remote access.
\end{takeout}
\section{Discussion}\label{sec:discussion} %

The preceding sections show that infostealer infections rarely stand alone. A single compromise exposes
work and personal accounts together, most of them protected by a handful of reused passwords, and supplies
credentials that resurface in phishing and ransomware campaigns. These findings hold only within the limits of
our methods.

\subsection{Limitations}\label{sec:limitations}

\textbf{Sample bias:}
This study does not represent a sample of all infostealer infections or of the broader internet population;
it is based on criminal actors sharing publicly available logs, collected in January 2025 and covering victim
device data from 2024 to 2025.
As a result, we underestimate the extent of infostealer attacks and miss very targeted ones.
Logs with the highest resale value are the least likely to be given away, so exposure among critical
organizations is probably worse than we measure.

\noindent \textbf{Source coverage:}
Our coverage is uneven by design. The collection plan emphasizes underground forums and open Telegram and Discord
channels, which can under-represent closed ecosystems, non-indexed communities, and sources reachable only through
different languages or access modalities, a common constraint in Cybersecurity Threat Landscape (CTL) style
production~\citep{ENISA-CTL-2025}.

\noindent \textbf{Third-party sandbox output:}
The family characterization in \autoref{sec:tooling} rests on sandbox reports published by external
services. We did not control the execution environment, we did not validate the family labels or the
ATT\&CK mapping ourselves, and we cannot determine how many samples failed to produce a report, so the
counts in \autoref{tbl:ttps} are of labelled samples rather than of successful executions.

\noindent \textbf{Unknown infection vector:}
Victims heavily used online gaming services, which suggests a link to the infection vector, but we cannot
verify it. Our behavioral findings therefore describe online gaming consumers more than the general
population.

\noindent \textbf{Authenticated services only:}
We study only the services where a victim authenticated with credentials,
so we do not study the full browsing history and miss any site visited without
logging in.

\noindent \textbf{Unrecognized services:}
Many services cannot be identified from their URLs alone, so for those entries we often cannot determine which
account or service a victim's compromised credentials belong to.

\noindent \textbf{Asymmetric deduplication:}
We distinguish Windows victims by machine identifier and macOS victims by IP address, so a shared
address merges several macOS devices into a single record. This undercounts macOS victims and
inflates their per-victim rates. We bound both effects using the Windows victims, for whom we hold
both identifiers: deduplicating a random Windows subsample the size and country mix of the macOS
population by address alone hides 0.9\,\% [0.6--1.2] of its devices and leaves 0.6\,\% [0.4--0.8] of
its records holding more than one device, over 200 draws with 95\,\% percentile intervals.
The 4,523 macOS records therefore stand for roughly 4,563 devices [4,551--4,577]. Standardizing on the
number of stored domains absorbs much of the remaining effect, so the macOS rates
in~\autoref{tbl:os_strata} are inflated by well under one percent.
 
\subsection{Lessons learned}\label{sec:learned}

A compromised host exposes passwords, session tokens, and cryptographic keys alike, so controls built around secrets and signatures no longer suffice on their own;
protection has to move toward hardware-bound authentication and behavior-based detection.

\begin{lesson}
  Session hijacking via cookie theft enables attackers to bypass multi-factor authentication.
  Passkeys and other FIDO2-based mechanisms strengthen authentication but do not protect against session replay attacks:
  an adversary who exfiltrates authenticated session artifacts (e.g., infostealer malware copying web browser cookies) can reuse them to impersonate the user.
  Mitigating this class of attacks requires the adoption of device-bound session mechanisms:
  Chrome (Windows) is rolling out Device Bound Session Credentials (DBSC)\footurl{https://developer.chrome.com/docs/web-platform/device-bound-session-credentials},
  where session tokens are cryptographically bound to a hardware-backed key (e.g., Trusted Platform Module [TPM]).
  Protection against session hijacking requires the use of short-lived hardware-bound tokens:
  the (web) server must periodically ask the (web browser) client to provide cryptographic hardware-backed verification during session use.
\end{lesson}

By collecting private keys, certificates, and VPN configurations, infostealers hand attackers access to
organizations' internal networks. Because they harvest credentials across so many critical services,
password complexity policies, however necessary, cannot protect user accounts on their own.

\begin{lesson}
  Endpoint protection mechanisms must refine their strategies to better protect hosts from data exfiltration.
  \autoref{tbl:ttps} shows that all seven families read data from the local system (T1005) and
  credentials from files (T1552.001, or its superseded identifier T1081), and query system
  information (T1082, T1012), as core TTPs.
  Monitoring access to well-known browser files (e.g., Chrome's \texttt{Login Data} and \texttt{Cookies} SQLite databases,
  Firefox's \texttt{logins.json}) during runtime would create a high-fidelity detection signal applicable across all studied families,
  enabling endpoint protection to block exfiltration even when the malware binary is novel or obfuscated.
\end{lesson}

The same signal extends beyond browser files to the encryption keys, VPN settings, and
cryptocurrency wallets that infostealers harvest. Defenses also cannot assume a single target
platform.

\begin{lesson}
  The dataset contains 4,523 macOS infections.
  This finding challenges the common perception that macOS systems face negligible malware risk and underscores that macOS is not immune to infostealers.
\end{lesson}

The macOS victims differ from the Windows majority in what they expose rather than in how carelessly
they behave, holding more institutional, financial and workplace credentials, twice the rate of
internal hosts, and slightly fewer reused passwords (\autoref{sec:family_strata}). That profile fits
a work device better than a personal one, though our records do not say who owns the machine.
Establishing whether macOS victims are disproportionately reached through employer-issued hardware
is worth doing, because a managed fleet can be defended centrally and a personal laptop holding work
credentials cannot.

\section{Conclusion}\label{sec:conclusion}

To our knowledge, this work presents the first ethically guided, victim-centered analysis at this scale of openly shared logs
from modern infostealer malware,
covering 170,298 distinct victims worldwide compromised in 2024--2025. By applying data minimization and
privacy-preserving anonymization, we draw insights without exposing direct personally identifiable
information, recoverable credentials, or live access tokens.

Our findings demonstrate that infostealers constitute a pervasive and significant threat to both individuals and organizations.
The dataset reveals broad exposure across sensitive sectors, including government, military,
law enforcement, and critical infrastructure, underscoring the national security implications of infostealer compromise.
Compromised credentials span high-value targets such as academic institutions (all eight Ivy League universities),
international organizations (\texttt{interpol.int}, n = 2), and government agencies (\texttt{dhs.gov}, n = 42).

Our analysis highlights the central role of human factors: widespread credential reuse
(\rounding{151547/170298*100}\,\% of victims), poor digital hygiene, and susceptibility to revictimization through phishing
and ransomware. These findings reinforce that technical defenses alone are insufficient without
user-centered security practices, awareness training, and periodic hardware-backed authentication
through short-lived session tokens.

Beyond the empirical findings, this work demonstrates that illicit cybercrime datasets can be transformed
into privacy-preserving research artifacts suitable for reproducible scientific analysis.
We share the anonymized dataset and the complete analysis code under restricted access, granted on request
for academic use (\refavail{}), and encourage the security community to build upon them.
In particular, future work can examine how victim behavior relates to their exposure.

\ifELSEVIER{}
\ifELSEVIER{}
\section*{CRediT authorship contribution statement}
\else
\section{Contributions}\label{sec:contributions}
\fi

Contributor Roles Taxonomy (\href{https://credit.niso.org/}{CRediT}) statement of authorship contribution.

\begin{table}[!htb]
  \rowcolors{2}{gray!10}{white}
\iftrue%
  \resizebox{0.98\linewidth}{!}{%
    \begin{tabular}{@{}l*{10}{c}@{}}
      \toprule
      \textbf{Contributions} & \textbf{A.P.} & \textbf{J.N.} & \textbf{D.A.} & \textbf{S.M.} & \textbf{F.C.} & \textbf{M.N.} & \textbf{J.I.} & \textbf{J.S.} & \textbf{C.P.} & \textbf{G.S.} \\
      \midrule
Conceptualization & $\bullet$ & $\bullet$ & $\bullet$ & $\circ$ & $\circ$ & $\bullet$ & $\bullet$ & $\circ$ & $\bullet$ & $\bullet$ \\
Data curation & $\bullet$ & $\bullet$ & $\circ$ & $\bullet$ & $\circ$ & $\circ$ & $\circ$ & $\circ$ & $\bullet$ & $\circ$ \\
Formal Analysis & $\bullet$ & $\bullet$ & $\circ$ & $\bullet$ & $\circ$ & $\circ$ & $\circ$ & $\circ$ & $\bullet$ & $\circ$ \\
Funding acquisition & $\circ$ & $\bullet$ & $\bullet$ & $\circ$ & $\bullet$ & $\bullet$ & $\circ$ & $\circ$ & $\bullet$ & $\bullet$ \\
Investigation & $\bullet$ & $\bullet$ & $\bullet$ & $\bullet$ & $\bullet$ & $\circ$ & $\bullet$ & $\circ$ & $\bullet$ & $\bullet$ \\
Methodology & $\bullet$ & $\bullet$ & $\bullet$ & $\circ$ & $\bullet$ & $\bullet$ & $\bullet$ & $\circ$ & $\bullet$ & $\bullet$ \\
Project administration & $\circ$ & $\bullet$ & $\bullet$ & $\circ$ & $\bullet$ & $\circ$ & $\bullet$ & $\circ$ & $\bullet$ & $\bullet$ \\
Resources & $\circ$ & $\bullet$ & $\bullet$ & $\circ$ & $\circ$ & $\bullet$ & $\circ$ & $\circ$ & $\bullet$ & $\bullet$ \\
Software & $\bullet$ & $\bullet$ & $\circ$ & $\bullet$ & $\circ$ & $\circ$ & $\circ$ & $\circ$ & $\bullet$ & $\circ$ \\
Supervision & $\circ$ & $\bullet$ & $\circ$ & $\circ$ & $\circ$ & $\circ$ & $\circ$ & $\circ$ & $\bullet$ & $\bullet$ \\
Validation & $\bullet$ & $\circ$ & $\circ$ & $\circ$ & $\circ$ & $\circ$ & $\circ$ & $\circ$ & $\circ$ & $\circ$ \\
Visualization & $\bullet$ & $\bullet$ & $\circ$ & $\bullet$ & $\circ$ & $\circ$ & $\circ$ & $\circ$ & $\bullet$ & $\circ$ \\
Writing---original draft & $\bullet$ & $\bullet$ & $\bullet$ & $\circ$ & $\bullet$ & $\circ$ & $\bullet$ & $\circ$ & $\bullet$ & $\bullet$ \\
Writing---review \& editing & $\bullet$ & $\bullet$ & $\bullet$ & $\bullet$ & $\bullet$ & $\bullet$ & $\circ$ & $\bullet$ & $\bullet$ & $\bullet$ \\
      \bottomrule
      \end{tabular}%
}
\else
  \Huge\textbf{PLACE\\HOLDER}
\fi
\end{table}

\FloatBarrier{}
\ifANON\else
\ifELSEVIER{}

\section*{Declaration of competing interest}

The authors declare that they have no known competing financial interests or
personal relationships that could have appeared to influence the work reported
in this paper.

\section*{Acknowledgments}\label{sec:acknowledgments}

This work was supported by the European Commission under the Horizon Europe funding programme, as part of the project SafeHorizon (Grant Agreement number 101168562). %
This work was partially supported by Ministerio de Ciencia, Innovaci\'{o}n y Universidades, Gobierno de Espa\~na
(Agencia Estatal de Investigaci\'{o}n, Fondo Europeo de Desarrollo Regional --FEDER--, European Union)
under the research grant PID2021-127409OB-C33 CONDOR, and by AGAUR with the project ASCLEPIUS (2021SGR-00111), and by Comunidad Aut\'{o}noma de Madrid, CIRMA-CAM Project (TEC-2024/COM-404).
Fran Casino was supported by the Spanish Ministry of Science and Innovation under the \qt{Ram\'{o}n y Cajal} programme (RYC2023-044857-I). %

The authors would like to thank Sara Degli-Esposti for her valuable feedback and support. %

Views and opinions expressed are however those of the author(s) %
only and do not necessarily reflect those of the European Union.
Neither the European Union nor the granting authority can be held responsible for them.

The funding sources had no involvement in the study design; in the collection,
analysis, and interpretation of data; in the writing of the report; or in the
decision to submit the article for publication.

\else

\section*{Acknowledgments}\label{sec:acknowledgments}

\begin{acks}

This work was supported by the European Commission under the Horizon Europe funding programme, as part of the project SafeHorizon (Grant Agreement number 101168562). %
This work was partially supported by Ministerio de Ciencia, Innovaci\'{o}n y Universidades, Gobierno de Espa\~na
(Agencia Estatal de Investigaci\'{o}n, Fondo Europeo de Desarrollo Regional --FEDER--, European Union)
under the research grant PID2021-127409OB-C33 CONDOR, and by AGAUR with the project ASCLEPIUS (2021SGR-00111), and by Comunidad Aut\'{o}noma de Madrid, CIRMA-CAM Project (TEC-2024/COM-404).
Fran Casino was supported by the Spanish Ministry of Science and Innovation under the \qt{Ram\'{o}n y Cajal} programme (RYC2023-044857-I). %

The authors would like to thank Sara Degli-Esposti for her valuable feedback and support. %

Views and opinions expressed are however those of the author(s) %
only and do not necessarily reflect those of the European Union.
Neither the European Union nor the granting authority can be held responsible for them.

The authors declare no competing interests.

\end{acks}
\fi
\fi
\ifELSEVIER{}
\section*{Declaration of generative AI and AI-assisted technologies in the manuscript preparation process}

During the preparation of this work the authors used QuillBot and Claude Opus in
order to check grammar and proofread the text, and ChatGPT in order to improve
the Python analysis code and LaTeX markup. No AI-generated text was included in
the manuscript; the authors wrote all the text. After using these tools/services,
the authors reviewed and edited the content as needed and take full responsibility
for the content of the published article.
\else
\section{Use of AI tools}\label{sec:ml}

No AI-generated text was included in the manuscript; the authors wrote all the text.
We used \url{https://quillbot.com/} online Grammar Checker and Claude Opus to proofread our article.
We used ChatGPT (\url{https://chat.openai.com/}) to improve Python code and LaTeX markup.
\fi
\ifELSEVIER{}
\section*{Data availability}
\phantomsection\label{sec:availability}
\else
\section{Open Science}\label{sec:availability}
\fi

In accordance with the data and open-science policy of the journal,
we provide the artifacts
necessary to replicate our calculations, analyses, tables, and
figures. %
The dataset of infostealer malware victims (processed and anonymized) and our analysis code are
deposited on Zenodo under restricted access~\citep{zenodo_dataset}.
Access is granted on request to researchers who state an academic purpose,
following recent measurement studies that share sensitive data on the same terms~\citep{yaben2026measuring,szakaly2025current,marjanov2026stayin}.
A request states the applicant's institutional affiliation and academic role and describes the
intended research use. Requesting access constitutes acceptance of the terms of use, which forbid
attempts to re-identify the victims, redistribution of the data to third parties, and any
operational or commercial use of the records. The corresponding author reviews each request against
those terms and replies within six weeks, usually sooner.
The deposit contains:
(i) the processed infostealer victim dataset with direct personally identifiable information removed or replaced,
(ii) the CSV files containing the phishing dataset,
(iii) the complete analysis scripts, and
(iv) a detailed \texttt{README.md} documenting the full workflow, including data preparation, analysis steps, and figure generation.

The documentation enables researchers granted access to fully reproduce our results and build upon our dataset for future studies,
while adhering to all legal and ethical constraints.
\fi

\ifELSEVIER{}
\bibliographystyle{elsarticle-num}
\fi

\appendix

\section*{Appendix}\label{app:appendix}

\renewcommand{\sectionautorefname}{Appendix}
\renewcommand{\thesection}{\Alph{section}}

\ifELSEVIER{}
\counterwithin{table}{section}
\counterwithin{figure}{section}
\fi

\ifELSEVIER\else
\ifELSEVIER{}
\section*{Data availability}
\phantomsection\label{sec:availability}
\else
\section{Open Science}\label{sec:availability}
\fi

In accordance with the data and open-science policy of the journal,
we provide the artifacts
necessary to replicate our calculations, analyses, tables, and
figures. %
The dataset of infostealer malware victims (processed and anonymized) and our analysis code are
deposited on Zenodo under restricted access~\citep{zenodo_dataset}.
Access is granted on request to researchers who state an academic purpose,
following recent measurement studies that share sensitive data on the same terms~\citep{yaben2026measuring,szakaly2025current,marjanov2026stayin}.
A request states the applicant's institutional affiliation and academic role and describes the
intended research use. Requesting access constitutes acceptance of the terms of use, which forbid
attempts to re-identify the victims, redistribution of the data to third parties, and any
operational or commercial use of the records. The corresponding author reviews each request against
those terms and replies within six weeks, usually sooner.
The deposit contains:
(i) the processed infostealer victim dataset with direct personally identifiable information removed or replaced,
(ii) the CSV files containing the phishing dataset,
(iii) the complete analysis scripts, and
(iv) a detailed \texttt{README.md} documenting the full workflow, including data preparation, analysis steps, and figure generation.

The documentation enables researchers granted access to fully reproduce our results and build upon our dataset for future studies,
while adhering to all legal and ethical constraints.
 \fi

\section{Ethical considerations}\label{sec:ethics}

\textit{The SafeHorizon project partner collected the raw data files in January 2025 as part of monitoring infostealer malware operations.
The partner's operational collection was conducted under its own legal and compliance framework for
threat-intelligence operations.
The SafeHorizon project applied for ethical approval for processing this data into a research dataset shared across the project.
The Ethics Advisory Board of the SafeHorizon project (101168562) %
formally approved the research methodologies and data processing protocols for the infostealer malware data on 26 August 2025.
They reviewed and endorsed the approach to data collection and dissemination, including
the release of the anonymized dataset on Zenodo for scholarly use, and
carefully considered the ethical implications of conducting research without direct consent from the affected individuals,
emphasizing strict adherence to data minimization principles.}

\textbf{Data minimization and anonymization.} Given that the dataset
of infostealer logs analyzed herein originated from illicit
activities, we followed the guidance of Thomas et
al.~\citep{DBLP:conf/imc/ThomasPHCB17} and recent discussions in the
security ethics literature~\citep{kohno} to take measures based on the
analysis of the potential harms and benefits of this research. To
minimize potential harms to victims and prevent re-identification, we
applied the data minimization principle, as defined in the EU General
Data Protection Regulation. An example of anonymized victim data is
included as a JSON\@ snapshot in \autoref{lst:json} in~\autoref{app:JSON}.
Personally Identifiable Information, such as
IP addresses, machine identifiers, usernames, and passwords, was
replaced with a salted hash during data
preprocessing. Further procedures were applied to reduce the risk of
accidentally encountering additional PII in the information leaked
from the compromised services. For example, to avoid having usernames
corresponding to real family names, all were replaced.

\textbf{Removal of raw data.}
To protect the victims, we permanently removed the original raw data after processing it into a research database
that we can share with the research community.
Releasing the anonymized dataset and our analysis code enables the community to reproduce our findings and study how to
protect users and secure systems against infostealers, without processing the original sensitive logs. This keeps
our work on an ethically constructed dataset with a balanced risk-benefit ratio.

\textbf{Controlled access.} The dataset is released through Zenodo under restricted access, granted
on request to researchers who state an academic purpose and accept a data use agreement.
Privacy in software logs is still an underexplored problem with no settled definition of what counts
as sensitive, and combinations of quasi-identifiers such as operating system, language, VPN status,
and autonomous system can support re-identification even after direct identifiers are
removed~\citep{aghili2025protecting}.
We therefore treat research data as open as possible and as closed as necessary, and
pair the field-level generalization in~\autoref{tbl:field_treatment} with access control rather than
relying on either alone.
Recent measurement studies that work with sensitive network and infrastructure data share it on the
same terms~\citep{yaben2026measuring,szakaly2025current,marjanov2026stayin}.

\textbf{Respect for persons.} From a deontological perspective,
victims shall be protected from further harm. We honored this duty
through data minimization, anonymization, removal of
residual identifiers, and exclusion of any content that could be
operationally misused.
Tangible harms such as re-compromise were mitigated
through anonymization, salted hashing, and avoiding host contact
(e.g., service status was assessed via publicly available tools such
as Shodan).
Residual re-identification risk is limited to potential
inferences from external datasets, which we consider unlikely given
applied protections. Following recommendations originating from both
consequentialist and deontological approaches~\citep{kohno}, we
prioritize the defense of the victims through the deontological duty
to privacy protection. In practice, we reverse-engineered infostealer malware to identify exfiltration endpoints and collected samples
without contacting live victims, so the analysis never required handling victims' PII directly. The released dataset does not store
recoverable credential material, private keys, or live access tokens.
This minimizes the risk that researchers are exposed to information
that could be exploited to target or attack a person or an
organization.

\textbf{Researcher well-being.}  The team was not exposed to harmful
or illicit material: web-content categories, including highly
sensitive ones such as child sexual abuse material, were assigned
automatically by the third-party Cisco Talos URL Reputation API from
URL strings, never by human inspection of content.
Access to sensitive data was restricted, encrypted, and limited to
necessary analysis.

\textbf{Responsible disclosure.}
During the course of this research, we identified active illegal websites, malware distribution infrastructure, and other security threats.
We reported these findings to the appropriate authorities and trusted operational partners.
Through the SafeHorizon project, partners coordinate responses with national Computer Emergency Response Teams (CERTs),
law enforcement agencies, and other relevant stakeholders to support the mitigation.
Individual victims were not contacted directly,
as the released research dataset contains only anonymized information and does not permit re-identification.

\textbf{Risk-benefit analysis.} Following the Menlo
Report~\citep{menlo} and consequentialist reasoning, we argue that the
benefits of understanding modern infostealer operations, exposure
patterns, and hygiene failures, summarized in the lessons learned
section of this paper, outweigh the residual privacy risk. Specific stakeholders who can benefit from our
research include: (i) individuals whose devices were compromised, who
can learn how to better protect themselves; (ii) organizations whose
members had their credentials compromised and should revise their
security awareness training and procedures; (iii) law enforcement
agents and incident response teams who can identify groups of users or
services more at risk and take measures to prevent future harm; and
(iv) security practitioners and scholars who can replicate the study
advancing further the debate about the trade-off between data
minimization and data usability. Finally, we believe that this paper
does not disclose recoverable credentials, live access tokens, or
operational details that would enable malevolent actions.
 
\ifELSEVIER\else
\ifELSEVIER{}
\section*{Declaration of generative AI and AI-assisted technologies in the manuscript preparation process}

During the preparation of this work the authors used QuillBot and Claude Opus in
order to check grammar and proofread the text, and ChatGPT in order to improve
the Python analysis code and LaTeX markup. No AI-generated text was included in
the manuscript; the authors wrote all the text. After using these tools/services,
the authors reviewed and edited the content as needed and take full responsibility
for the content of the published article.
\else
\section{Use of AI tools}\label{sec:ml}

No AI-generated text was included in the manuscript; the authors wrote all the text.
We used \url{https://quillbot.com/} online Grammar Checker and Claude Opus to proofread our article.
We used ChatGPT (\url{https://chat.openai.com/}) to improve Python code and LaTeX markup.
\fi
 
\ifANON\else
\ifELSEVIER{}
\section*{CRediT authorship contribution statement}
\else
\section{Contributions}\label{sec:contributions}
\fi

Contributor Roles Taxonomy (\href{https://credit.niso.org/}{CRediT}) statement of authorship contribution.

\begin{table}[!htb]
  \rowcolors{2}{gray!10}{white}
\iftrue%
  \resizebox{0.98\linewidth}{!}{%
    \begin{tabular}{@{}l*{10}{c}@{}}
      \toprule
      \textbf{Contributions} & \textbf{A.P.} & \textbf{J.N.} & \textbf{D.A.} & \textbf{S.M.} & \textbf{F.C.} & \textbf{M.N.} & \textbf{J.I.} & \textbf{J.S.} & \textbf{C.P.} & \textbf{G.S.} \\
      \midrule
Conceptualization & $\bullet$ & $\bullet$ & $\bullet$ & $\circ$ & $\circ$ & $\bullet$ & $\bullet$ & $\circ$ & $\bullet$ & $\bullet$ \\
Data curation & $\bullet$ & $\bullet$ & $\circ$ & $\bullet$ & $\circ$ & $\circ$ & $\circ$ & $\circ$ & $\bullet$ & $\circ$ \\
Formal Analysis & $\bullet$ & $\bullet$ & $\circ$ & $\bullet$ & $\circ$ & $\circ$ & $\circ$ & $\circ$ & $\bullet$ & $\circ$ \\
Funding acquisition & $\circ$ & $\bullet$ & $\bullet$ & $\circ$ & $\bullet$ & $\bullet$ & $\circ$ & $\circ$ & $\bullet$ & $\bullet$ \\
Investigation & $\bullet$ & $\bullet$ & $\bullet$ & $\bullet$ & $\bullet$ & $\circ$ & $\bullet$ & $\circ$ & $\bullet$ & $\bullet$ \\
Methodology & $\bullet$ & $\bullet$ & $\bullet$ & $\circ$ & $\bullet$ & $\bullet$ & $\bullet$ & $\circ$ & $\bullet$ & $\bullet$ \\
Project administration & $\circ$ & $\bullet$ & $\bullet$ & $\circ$ & $\bullet$ & $\circ$ & $\bullet$ & $\circ$ & $\bullet$ & $\bullet$ \\
Resources & $\circ$ & $\bullet$ & $\bullet$ & $\circ$ & $\circ$ & $\bullet$ & $\circ$ & $\circ$ & $\bullet$ & $\bullet$ \\
Software & $\bullet$ & $\bullet$ & $\circ$ & $\bullet$ & $\circ$ & $\circ$ & $\circ$ & $\circ$ & $\bullet$ & $\circ$ \\
Supervision & $\circ$ & $\bullet$ & $\circ$ & $\circ$ & $\circ$ & $\circ$ & $\circ$ & $\circ$ & $\bullet$ & $\bullet$ \\
Validation & $\bullet$ & $\circ$ & $\circ$ & $\circ$ & $\circ$ & $\circ$ & $\circ$ & $\circ$ & $\circ$ & $\circ$ \\
Visualization & $\bullet$ & $\bullet$ & $\circ$ & $\bullet$ & $\circ$ & $\circ$ & $\circ$ & $\circ$ & $\bullet$ & $\circ$ \\
Writing---original draft & $\bullet$ & $\bullet$ & $\bullet$ & $\circ$ & $\bullet$ & $\circ$ & $\bullet$ & $\circ$ & $\bullet$ & $\bullet$ \\
Writing---review \& editing & $\bullet$ & $\bullet$ & $\bullet$ & $\bullet$ & $\bullet$ & $\bullet$ & $\circ$ & $\bullet$ & $\bullet$ & $\bullet$ \\
      \bottomrule
      \end{tabular}%
}
\else
  \Huge\textbf{PLACE\\HOLDER}
\fi
\end{table}

\FloatBarrier{}
\ifELSEVIER{}

\section*{Declaration of competing interest}

The authors declare that they have no known competing financial interests or
personal relationships that could have appeared to influence the work reported
in this paper.

\section*{Acknowledgments}\label{sec:acknowledgments}

This work was supported by the European Commission under the Horizon Europe funding programme, as part of the project SafeHorizon (Grant Agreement number 101168562). %
This work was partially supported by Ministerio de Ciencia, Innovaci\'{o}n y Universidades, Gobierno de Espa\~na
(Agencia Estatal de Investigaci\'{o}n, Fondo Europeo de Desarrollo Regional --FEDER--, European Union)
under the research grant PID2021-127409OB-C33 CONDOR, and by AGAUR with the project ASCLEPIUS (2021SGR-00111), and by Comunidad Aut\'{o}noma de Madrid, CIRMA-CAM Project (TEC-2024/COM-404).
Fran Casino was supported by the Spanish Ministry of Science and Innovation under the \qt{Ram\'{o}n y Cajal} programme (RYC2023-044857-I). %

The authors would like to thank Sara Degli-Esposti for her valuable feedback and support. %

Views and opinions expressed are however those of the author(s) %
only and do not necessarily reflect those of the European Union.
Neither the European Union nor the granting authority can be held responsible for them.

The funding sources had no involvement in the study design; in the collection,
analysis, and interpretation of data; in the writing of the report; or in the
decision to submit the article for publication.

\else

\section*{Acknowledgments}\label{sec:acknowledgments}

\begin{acks}

This work was supported by the European Commission under the Horizon Europe funding programme, as part of the project SafeHorizon (Grant Agreement number 101168562). %
This work was partially supported by Ministerio de Ciencia, Innovaci\'{o}n y Universidades, Gobierno de Espa\~na
(Agencia Estatal de Investigaci\'{o}n, Fondo Europeo de Desarrollo Regional --FEDER--, European Union)
under the research grant PID2021-127409OB-C33 CONDOR, and by AGAUR with the project ASCLEPIUS (2021SGR-00111), and by Comunidad Aut\'{o}noma de Madrid, CIRMA-CAM Project (TEC-2024/COM-404).
Fran Casino was supported by the Spanish Ministry of Science and Innovation under the \qt{Ram\'{o}n y Cajal} programme (RYC2023-044857-I). %

The authors would like to thank Sara Degli-Esposti for her valuable feedback and support. %

Views and opinions expressed are however those of the author(s) %
only and do not necessarily reflect those of the European Union.
Neither the European Union nor the granting authority can be held responsible for them.

The authors declare no competing interests.

\end{acks}
\fi
 \fi
\fi

\section{Victim statistics}\label{sec:stats}

\autoref{tbl:victim_statics} gives the most common (a) languages and (b) autonomous systems of the
infostealer victims. The \qt{Other} and \qt{Aggregated} rows hold the values that the generalization
described in~\autoref{sec:datacollection} replaced, so the named rows do not cover every victim.

\begin{table}[!htb]
  \centering
  \caption{Victim statistics.}\label{tbl:victim_statics}
  \scriptsize
  \begin{subtable}[h]{0.35\columnwidth}
  \caption{Victims' languages.}\label{tbl:language}
\rowcolors{2}{gray!10}{white}
  \begin{tabular}{lr}
    \toprule
    \textbf{Language} & \textbf{Count}\\
    \midrule
en & 86,495 \\
es & 21,993 \\
pt & 15,789 \\
fr & 7,309 \\
N/A & 6,293 \\
tr & 4,903\\
ar & 3,944\\
th & 3,452\\
Other & 2,142\\
pl & 2,100\\
\bottomrule
  \end{tabular}
\end{subtable}
\hfill
\begin{subtable}[h]{0.64\columnwidth}
  \centering
  \caption{ASN used by the victims.}\label{tbl:asn}
\rowcolors{2}{gray!10}{white}
  \begin{tabular}{lr}
    \toprule
    \textbf{ASN} & \textbf{Count}  \\
    \midrule
Aggregated & 61,385\\
Reliance Jio Infocomm Limited & 5,345\\
TE Data & 3,739\\
PT Telekomunikasi Indonesia & 3,045\\
Bharti Airtel Ltd.\ Telemedia & 2,885\\
VNPT Corp & 2,811\\
Viettel Group & 2,597\\
COMCAST-7922 & 2,346\\
Telecom Algeria & 2,122\\
Philippine telecom PLDT, Inc. & 2,093\\
Claro NXT Telecomunicacoes & 1,896\\
\bottomrule
  \end{tabular}
\end{subtable}
\end{table}

\section{Infostealer offerings}\label{app:offerings}

\begin{figure}[!htb]
  \centering
  \includegraphics[width=.8\columnwidth]{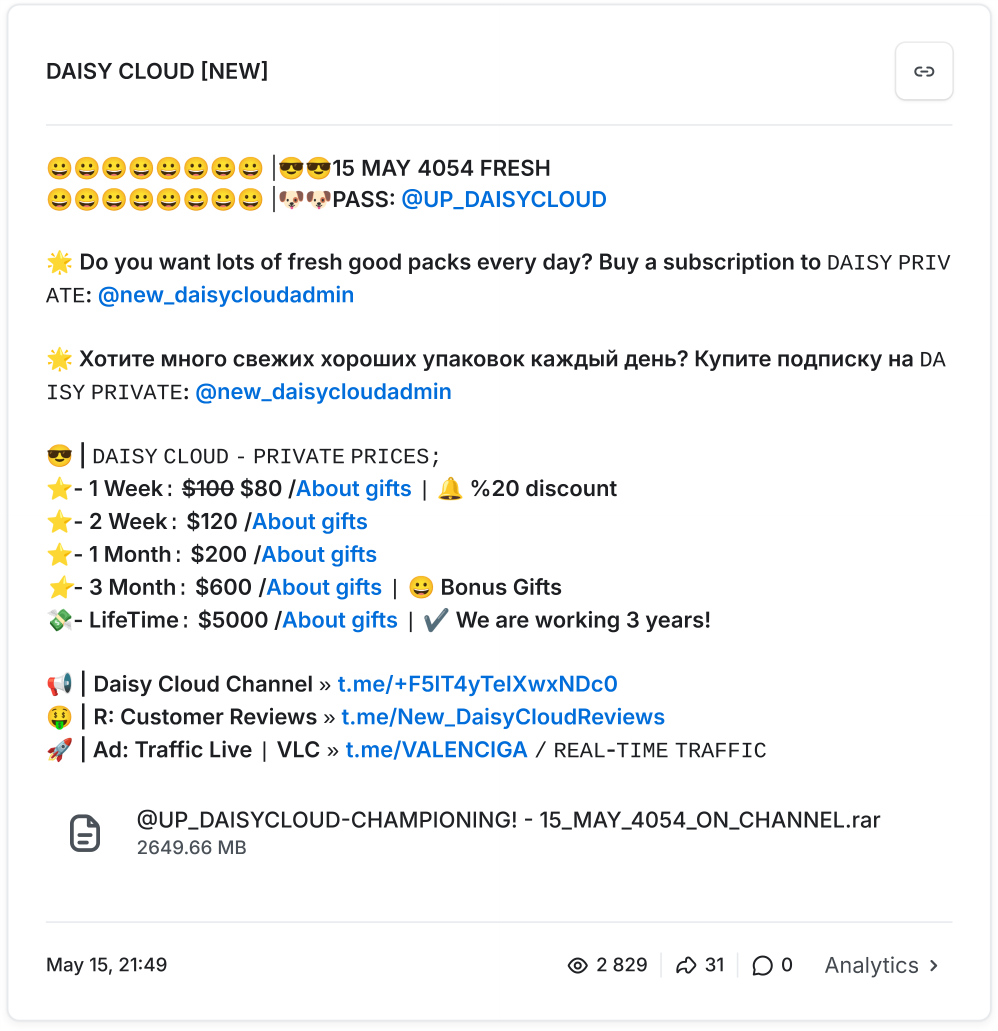}
  \caption{Infostealer malware logs offerings.}\label{fig:adv}
\end{figure}

\autoref{fig:adv} shows an infostealer operator's offerings on Telegram, using discounts and tiered pricing to
attract customers with different budgets.

\section{Example of infostealer data}\label{app:JSON}
\begin{lstlisting}[language=json,label={lst:json}, caption={A victim's profile from our dataset (0001021.json)}]
{
    "timestamp": "2024-12-30",
    "ip_address": "d17f353166",
    "ip_location": "US",
    "ip_asn": "ASN: 212238, ORG: Datacamp Limited",
    "is_vpn": "VPN",
    "location": "",
    "country": "US",
    "os": "Windows 11 Home",
    "machineid": "57d7535e8a",
    "language": "en",
    "keyboard_layouts": [],
    "passwords": ["a809434c42", "8c32f0d2f0", ...],
    "usernames": ["b8f2935589", "b8f2935589", ...],
    "urls": [
        "https://spotify.com",
        "https://uploadhaven.com",
        ...
    ],
    "onion_domain_names": [],
    "urls_count": 276,
    ...
}
\end{lstlisting}

\autoref{lst:json} is an example of anonymized victim data that is
included as a JSON\@ snapshot in our dataset.
\autoref{tbl:field_treatment} gives the treatment applied to each field.

As described in~\autoref{sec:ethics}, the \breaktt{ip_address}, \breaktt{machineid},
\breaktt{usernames}, and \breaktt{passwords} fields are all replaced with salted hashes
(\autoref{sec:datacollection}).
The URLs field retains the registered domain of each service the victim authenticated to,
including the critical-infrastructure domains reported in~\autoref{tab:illustrative_examples}.

\section{Infostealer C2 server infrastructure}\label{app:c2common}

\begin{figure}[!htb]
  \centering
  \includegraphics[width=.8\columnwidth]{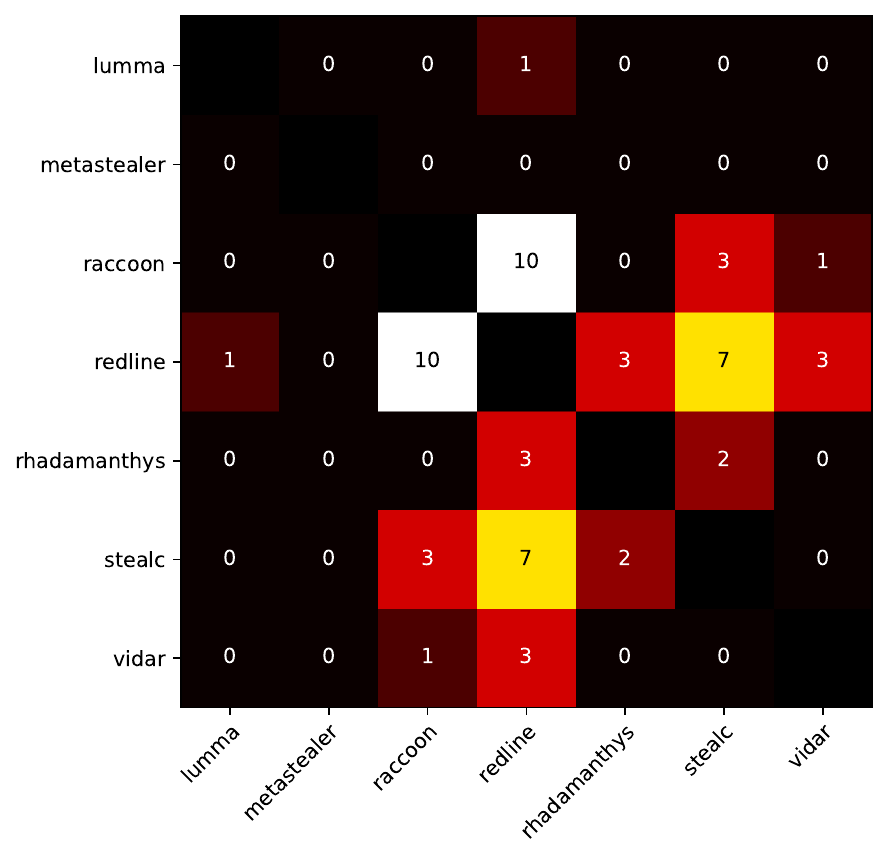}
  \caption{Common infrastructure used as C2 servers by different infostealers.}\label{fig:commoninfra}
\end{figure}

\autoref{fig:commoninfra} plots the C2 infrastructure shared between infostealers over our study
period. Non-zero values, most notably between RedLine and Raccoon (10 shared IPs) and RedLine and StealC
(7 shared IPs), show infrastructure overlap between families
({\renewcommand{\sectionautorefname}{Section}\autoref{sec:tooling}}).

\end{document}